\documentclass[sigconf,nonacm]{acmart}
\AtBeginDocument{%
  \providecommand\BibTeX{{%
    \normalfont B\kern-0.5em{\scshape i\kern-0.25em b}\kern-0.8em\TeX}}}

\setcopyright{acmlicensed}
\acmJournal{PACMMOD}
\acmYear{2024} \acmVolume{2} \acmNumber{1 (SIGMOD)} 
\acmArticle{53} \acmMonth{2} 
\acmDOI{10.1145/3639308}

\newcommand{\squishlist}{
   \begin{list}{$\bullet$}{%
        \setlength{\itemsep}{0pt}%
        \setlength{\parsep}{0pt}%
        \setlength{\topsep}{0pt}%
        \setlength{\partopsep}{0pt}%
        \setlength{\listparindent}{-2pt}%
        \setlength{\itemindent}{-5pt}%
        \setlength{\leftmargin}{1.2em}%
        \setlength{\labelwidth}{0em}%
        \setlength{\labelsep}{0.5em}%
    }
}

\newcommand{\squishend}{
    \end{list}  }

\usepackage[skip=5pt]{caption}
\usepackage{tablefootnote}
\usepackage{xcolor}
\usepackage{comment}
\usepackage{xspace}
\usepackage{tabularx}
\usepackage{tablefootnote}
\usepackage{threeparttable}

\usepackage{algorithm}

\usepackage{algpseudocode}
\algtext*{EndFor}% Remove "end while" text
\algtext*{EndIf}% Remove "end if" text

\usepackage[T1]{fontenc}
\usepackage{contour}
\usepackage{amsmath}
\usepackage{dsfont}
\usepackage{mfirstuc}
\usepackage{multirow}
\usepackage{listings}
\usepackage{float}
\usepackage{subcaption}

\newcommand{\yuanyuan}[1]{\textcolor{red}{\bf [Yuanyuan: #1]}}
\newcommand{\rana}[1]{\textcolor{cyan}{\bf [Rana: #1]}}
\newcommand{\hanxian}[1]{\textcolor{orange}{\bf [Hanxian: #1]}}
\newcommand{\tar}[1]{\textcolor{red}{\bf [Tarique: #1]}}
\newcommand{\todox}[1]{\textcolor{red}{\{\{#1\}\}}}
\newcommand {\sys}{\textsc{Sibyl}\xspace}
\newcommand {\sysModel}{\textsc{Sibyl-LSTMs}\xspace}
\newcommand{\cutNpack}{Cut-and-Pack\xspace}
\newcommand{\ie}{i.e.,~}
\newcommand{\eg}{e.g.,~}

\newcommand{\aka}{a.k.a.~}

\newcommand{\Eg}{E.g.,~} 

\newcommand{\revisioncolor}{black}
\newcommand{\revision}[1]{{\color{\revisioncolor} #1}}

\newcommand{\papertext}[1]{}

\newcommand{\techreporttext}[1]{}

\newcommand{\mssales}{{Sales}\xspace} 
\newcommand{\wmc}{{Telemetry}\xspace} 
\newcommand{\scope}{{SCOPE}\xspace} 
\newcommand{\microsoft}{{Microsoft}\xspace} 

\definecolor{codepurple}{HTML}{C42043}
\lstdefinestyle{sqlstyle}{
basicstyle=\footnotesize\ttfamily,
breaklines=true,
keywordstyle=\bfseries\color{codepurple},
language=sql
}
\lstnewenvironment{sqllisting}
    {\lstset{language=sql,escapeinside={(*@}{@*)}}}
    {}

\newcommand{\viewpoint}[2]{
\vspace{0.1cm}
\noindent\textbf{#1:} \textit{#2} %\\
}
\newtheorem{definition}{Definition}[section]

\begin{document}

\title{\sys: Forecasting Time-Evolving Query Workloads}


\author{Hanxian Huang}
\orcid{0000-0001-6338-3289}
\affiliation{%
\institution{University of California San Diego}
\country{USA}
}
\email{hah008@ucsd.edu}

\author{Tarique Siddiqui}
\orcid{0009-0002-0866-7275}
\affiliation{%
\institution{Microsoft Research}
\country{USA}
}
\email{Tarique.Siddiqui@microsoft.com}

\author{Rana Alotaibi}
\orcid{0009-0005-0457-8429}
\affiliation{%
\institution{Microsoft Gray Systems Lab}
\country{USA}
}
\email{ranaalotaibi@microsoft.com}

\author{Carlo Curino}
\orcid{0000-0003-3712-7358}
\affiliation{%
\institution{Microsoft Gray Systems Lab}
\country{USA}
}
\email{Carlo.Curino@microsoft.com}

\author{Jyoti Leeka}
\orcid{0000-0003-2920-1431}
\affiliation{%
\institution{Microsoft}
\country{USA}
}
\email{Jyoti.Leeka@microsoft.com}

\author{Alekh Jindal}
\orcid{0000-0001-8844-8165}
\affiliation{%
\institution{SmartApps}
\country{USA}
}
\email{alekh@smart-apps.ai}

\author{Jishen Zhao}
\orcid{0000-0002-8766-0946}
\affiliation{%
\institution{University of California San Diego}
\country{USA}
}
\email{jzhao@ucsd.edu}

\author{Jes\'us Camacho-Rodr\'iguez}
\orcid{0009-0008-9151-6024}
\affiliation{%
\institution{Microsoft Gray Systems Lab}
\country{USA}
}
\email{jesusca@microsoft.com}

\author{Yuanyuan Tian}
\orcid{0000-0002-6835-8434}
\affiliation{%
\institution{Microsoft Gray Systems Lab}
\country{USA}
}
\email{yuanyuantian@microsoft.com}
\authorsaddresses{}
\renewcommand{\shortauthors}{Hanxian Huang et al.}


\begin{abstract}
Database systems often rely on historical query traces to perform workload-based performance tuning.
However, real production workloads are time-evolving, making historical queries ineffective for optimizing future workloads. 
To address this challenge, we propose \sys, an end-to-end machine learning-based framework that accurately forecasts a sequence of future queries, with the entire query statements, in various prediction windows. Drawing insights from real-workloads, we propose template-based featurization techniques and develop \revision{a stacked-LSTM with an encoder-decoder architecture}
%a novel model that combines stacked-LSTM with an encoder-decoder architecture 
for accurate forecasting of query workloads. We also develop techniques to improve forecasting accuracy over large prediction windows
and achieve high scalability over large workloads with high variability in arrival rates of queries. Finally, we propose techniques to handle workload drifts. 
Our %extensive 
evaluation on four real workloads demonstrates that \sys can forecast workloads with an $87.3\%$ median F1 score, \revision{and can result in $1.7\times$ and $1.3\times$ 
performance improvement when applied to materialized view selection and index selection 
applications, respectively.} 
\end{abstract}

\begin{comment}
\begin{CCSXML}
<ccs2012>
   <concept>
       <concept_id>10002951.10002952</concept_id>
       <concept_desc>Information systems~Data management systems</concept_desc>
       <concept_significance>500</concept_significance>
       </concept>
 </ccs2012>
\end{CCSXML}

\ccsdesc[500]{Information systems~Data management systems}

\keywords{Query workload forecasting, Time-series forecasting}%, Workload-based optimization}

\received{July 2023}
\received[revised]{October 2023}
\received[accepted]{November 2023} 

\end{comment}

\maketitle

%\vspace{-2mm}
\section{Introduction}
\label{sec:intor}

Workload-based optimization is a critical aspect of database management, which tunes a database management system (DBMS) to maximize its performance for a specific workload. 
%It is essential to ensure that database systems can efficiently handle the demands of modern applications. 
Consequently, a large number of performance tuning tools have been developed for workload-based optimization. 
While execution statistics may be sufficient for optimizing certain aspects of the system (e.g., buffer pool size), many optimizations require an understanding of semantics of the queries, necessitating a workload query trace as input.
%While workload metrics, such as resource utilization and runtime statistics, may be sufficient for optimizing certain aspects of the system with some of these tools, other workload optimization tools require an understanding of the semantics of queries in the workload, making a workload query trace necessary as input.
For instance, many commercial database products~\cite{sqlserver,db2,oracle,informix} support physical design tools 
%provided by many established commercial database products~\cite{sqlserver,db2,oracle,informix}, 
such as Microsoft's AutoAdminn~\cite{autoadmin} and IBM's DB2 design advisor~\cite{db2advisor}. 
%have been developed. 
These tools automatically recommend physical design features, such as indexes, materialized views, partitioning schemes of tables, and multidimensional clustering (MDC)~\cite{mdc} of tables for a given workload of queries. 
%Another related workload-based optimization technique is semantic caching~\cite{DBLP:conf/sigmod/StonebrakerJGP90,DBLP:conf/edbt/ChenR94a,DBLP:conf/vldb/DarFJST96,DBLP:journals/pvldb/DurnerCL21}, which involves caching frequently accessed query results based on the semantic meaning of the queries to accelerate repeated execution. 
In fact, many of these %workload-based optimization techniques 
form the foundational pieces of the self-tuning~\cite{selftune}, self-managing~\cite{selfmanage}, and self-driving~\cite{oracleselfdrive,Pavlocidr17} databases.

For workload-based optimization, the input workload plays a crucial role and needs to be a good representation of the expected workload. 
Traditionally, historical query traces have been used as input workloads  with the assumption that workloads are mostly \emph{static}. However, as we discuss in \S\ref{sec:motivation}, many real workloads exhibit highly recurring query structures with changing patterns in both their arrival intervals and data accesses.
For instance, query templates are often shared across users, teams, and applications, but may be customized with different parameter values to access varying data at different points in time. Consider a log analysis query that reports errors for different devices and error types: \texttt{"SELECT * FROM T WHERE deviceType = ? AND errorType = ? AND eventDate BETWEEN ? AND ?"}. Although the query template is recurring, the parameter values may be customized depending on the reporting needs, e.g., different types of devices may be analyzed on different days of the week, and the granularity of the time interval may switch from daily to weekly over weekends. Thus, optimization recommendations (\eg recommended views) based solely on historical queries may not be effective for such time-evolving workload patterns. To adapt to evolving workloads, existing workload-based optimization tools can be modified and enhanced to accommodate workload changes, or alternatively, applied \textit{without modification} by substituting the history workload with a workload that more accurately represents the anticipated query trace in the future. This motivates the need for forecasting future workload.

As depicted in Table~\ref{tab:comparison}, prior research works~\cite{ma2018query,meduri2021evaluation,abebe2022tiresias} have made efforts to forecast different aspects of future workload, but none have addressed the challenge of predicting  future query traces with precise \textit{query statements} in a future \textit{time window} for \textit{time-evolving} workloads. The Q-Learning approach proposed by Meduri et al.~\cite{meduri2021evaluation} takes the current query as input and predicts the next \textit{one} query by forecasting query fragments separately and then assembling them into a complete query statement. 
%However, predicting fragments individually can result in invalid queries, requiring identification and repair of errors. 
For predicting the literals (or parameters) used in the query statement, Q-Learning only forecasts a bin of possible values instead of the exact values. 
%Furthermore, this approach only considers the current query as input, neglecting the time-evolving characteristics of the workload from a time-series perspective. 
QueryBot 5000~\cite{ma2018query} and Tiresias~\cite{abebe2022tiresias} focus on forecasting the \textit{arrival rate} of future query workload by training on patterns from historical query arrival rates. %So, with QueryBot\_5000, 
Hence, 
%one can predict, in a coarser grain, what types of queries and how many of them are expected in the future. 
these techniques can only predict the types of queries and the number of them expected in the future in a coarse granularity.
Yet they %cannot 
do not generate %provide 
the specific query statements for those workload-based optimization tools that require understanding the query semantics, especially in the context of time-evolving workloads.
%like in Example~1.
%QueryBot\_5000 leverages ensemble learning~\cite{opitz1999popular} to average the results of linear regression and RNNs. The linear regression does not support incremental fine-tuning and necessitates retraining from scratch for workload shifts. 
%We summarize the comparison between \sys\ and these two most related works in Table~\ref{tab:comparison}. 

\Huge
\begin{table}[ht!]
  \caption{ \sys\ vs. three closely related works.}
\label{tab:comparison}
\resizebox{\linewidth}{!}{
\begin{threeparttable}
\begin{tabular}{lccc}
\hline
& \begin{tabular}[c]{@{}c@{}}QueryBot 5000~\cite{ma2018query}\\  Tiresias~\cite{abebe2022tiresias}\end{tabular}
    & Q-Learning~\cite{meduri2021evaluation}  & \sys\    \\ \hline
What to predict     & Query arrival rate & Next query         & \begin{tabular}[c]{@{}c@{}}Future queries\\ \& arrival time\end{tabular} \\ \hline
Methods & \begin{tabular}[c]{@{}c@{}}Hybrid-ensemble \\Learning\end{tabular} & \begin{tabular}[c]{@{}c@{}}RNNs \&\\ Q-learning\end{tabular} & \sysModel\ \\ \hline
Applications       & \multicolumn{1}{c}{Index selection}    & \begin{tabular}[c]{@{}c@{}}Query\\ recommendation\end{tabular} & Physical design tools     \\ \hline
\begin{tabular}[l]{@{}l@{}}Time-evolving \\forecasting\end{tabular} %\tar{rename this to Future Arrival Rates?}
&    \textcolor{green!70!black}{\checkmark}     &   \textcolor{red}{$\times$}    &     \textcolor{green!70!black}{\checkmark} \\ \hline
\begin{tabular}[l]{@{}l@{}}%Various \tar{Variable?} 
Variable prediction \\windows\end{tabular}   &    \textcolor{green!70!black}{\checkmark}     &   \textcolor{red}{ $\times$}    &     \textcolor{green!70!black}{\checkmark} \\ \hline
%\begin{tabular}[l]{@{}l@{}}Forecast the exact\\ future queries\end{tabular}     &   \textcolor{red}{$\times$}    &  \textcolor{red}{$\times$}    &    \textcolor{green!70!black}{\checkmark}     \\ \hline
\begin{tabular}[l]{@{}l@{}}Forecast future\\ query statement  \end{tabular}       &   \textcolor{red}{$\times$}    &  \begin{tabular}[c]{@{}c@{}}\textcolor{green!70!black}{\checkmark}\tnote{$\dag$} \end{tabular}
&    \textcolor{green!70!black}{\checkmark}     \\ \hline
%\begin{tabular}[l]{@{}l@{}}Well-formed\\ query statement\end{tabular}       &   \textcolor{red}{$\times$}    &  \textcolor{red}{$\times$}    &    \textcolor{green!70!black}{\checkmark}     \\ \hline
\begin{tabular}[l]{@{}l@{}}Forecast \emph{precise}\\ query parameter values\end{tabular}       &   \textcolor{red}{$\times$}    &  \textcolor{red}{$\times$}\tnote{$\dag$}   &    \textcolor{green!70!black}{\checkmark}     \\ \hline
%\begin{tabular}[l]{@{}l@{}}Efficient workload\\ representation\end{tabular}     &    \textcolor{green!70!black}{\checkmark}   &  \textcolor{red}{$\times$}    &    \textcolor{green!70!black}{\checkmark}     \\ \hline
%\begin{tabular}[l]{@{}l@{}}Incremental fine-\\tuning on drifts\end{tabular}      &   \textcolor{red}{$\times$}    &    \textcolor{green!70!black}{\checkmark}&     \textcolor{green!70!black}{ \checkmark} \\ \hline
\end{tabular}

 \begin{tablenotes}
    %\footnotesize
    \huge
    \item[$\dag$] Meduri et al.~\cite{meduri2021evaluation} forecast future query statement with coarse-grained parameter value ranges.
  \end{tablenotes}

\end{threeparttable}
 
 }

\end{table}

\normalsize

To address these limitations, we propose \sys, an end-to-end machine learning (ML)-based workload forecasting framework, which can accurately predict the query statements in a future time window for time-evolving workload. 
Different from a specific workload-based optimization technique, by addressing this broader but also more challenging workload forecasting problem, our aim is to enable a wide range of existing workload-based optimization tools that were originally designed for static workloads to be directly applied \textit{without modification} for time-evolving workloads. %, as shown in Figure~\ref{fig:sibyl}.
We next highlight the major contributions of \sys. %\yuanyuan{I think we can consider removing figure 1. We now only have one application, plus 1.7X is not a very impressive number to highlight in the intro. }
\begin{comment}
\vspace{-3mm}
\begin{figure}[ht!]
   \centering
   \includegraphics[width=\linewidth]{figure/sibyl_1app.pdf}
   \caption{ \sys\ forecastes future workload accurately with an
$87.3\%$ median F1 score and improves workload-based optimization (\ie view selection by $1.7\times$) with the forecasted queries as input compared to conventional historical workload input.
   %, compared to using historical queries as input.
   %\yuanyuan{@Hanxian, can you add three dots between view selection and query result caching, to indicate that there are other applications that \sys can support. Also, I think color red is usually used for alarming things. Maybe we change the color red to color blue or green?}
   }
   \label{fig:sibyl}
   \vspace{-2mm}
 \end{figure}
\end{comment}

%\vspace{-2mm}
\subsection{\sys\ Contributions}

We motivate the \sys design %based on 
by qualitative and quantitative studies on four different real workloads (\S\ref{sec:motivation}). We identify three %crucial 
insights from our observations that are shared by many existing studies~\cite{jindal2021production,alibaba,ma2018query,abebe2022tiresias,CrocodileDB,p-store}: 1) real workloads are highly recurrent (sharing the same \textit{query templates} but with different \textit{query parameters}) as discussed earlier; 2) these recurrent queries often exhibit time-evolving behavior; and 3) they are highly predictable. Based on these insights, we first formalize a %simple 
next-$k$ workload forecasting problem that predicts the next $k$ queries in the future. We develop an ML-based technique %\sys\ 
that %seamlessly 
integrates both query arrival time and query parameters into a common framework, and forecasts the \emph{entire} statements of future recurrent queries and their arrival time. %Specifically,
To this end, we develop a template-based featurization method that captures a large number of parametric 
expressions (\S\ref{sec:featureEngineer}), and combine stacked LSTM~\cite{hochreiter1997long} with an encoder-decoder architecture~\cite{cho2014learning}, resulting in a model that we refer to as \sysModel (\S\ref{sec:templatemodels}), to better capture both the temporal dependencies and the possible inter-dependencies among query parameters, for \textit{per-template}, multi-variate, multi-step, time-series prediction.

We further identify a set of practical challenges in order to make the predictions usable for performance tuning tools. 
%However, there are some practical challenges that we need to address to make the predictions usable within performance tuning tools.
%For example, 
First, it is difficult to pre-determine the right number of queries ($k$) to forecast, to produce a useful workload for database optimization tools. Rather, a more practical version of the problem is to forecast the future queries for the next time interval $\Delta t$. But the number of queries expected in next $\Delta t$ varies substantially across templates.
%The template size (i.e., the expected number of queries for the template in next $\Delta t$ time interval) typically vary in real workloads. %Furthermore, 
Second, with one model per template, we need to limit the number of models to ensure the scalability of our design. To address these two issues, we extend the solution for the next-$k$ problem to the next-$\Delta t$ forecasting problem (\S\ref{sec:design-delta}). For templates with large expected numbers of queries, we `cut' them into sub-templates in order to employ the predicted next-$k$ queries to produce results for the next-$\Delta t$ problem. To reduce the number of models, we %also 
`pack' templates with a small expected number of queries into bins. We then build per-bin models (\S\ref{sec:perbinmodel}). %This
Our approach not only %provides 
yields accurate results for the next-$\Delta t$ forecasting problem, but also reduces the number of \sysModel\ models by up to $23\times$, resulting in a significant reduction in both time and storage overhead by up to $13.6\times$ and $6\times$, respectively. 

%Another 
A third %practical 
challenge is that real-world workloads change dynamically, i.e., new templates may emerge, while old templates may become less relevant. The evolving patterns of literals may also change. To capture such changes, %within our framework, 
\sys\ %provides 
adopts a feedback loop to handle workload shifts
(\S\ref{sec:feedbackloop}). %It
The feedback loop uses incremental learning to adapt the pre-trained \sysModel\ to the shifted workloads, achieving comparable accuracy to full training, with %only a few seconds of 
a negligible fine-tuning overhead.

%Overall, 
We integrate the forecasting model with the feedback loop to develop the end-to-end forecasting solution of \sys. The paper also presents a common effectiveness measurement that can be utilized for many performance tuning tools (\S\ref{sec:accuracy}).
We evaluate \sys\ on four real workloads (\S\ref{sec:eval}) and demonstrate its accurate forecasting ability. 
\revision{Furthermore, we apply \sys to two database applications on real workloads: materialized view  and index selection, and our results show $1.7\times$ and $1.3\times$ improvement, respectively, using forecasted workloads compared to historical workloads.}

%Furthermore, we apply \sys\ to %two database applications on real workloads: 
%the materialized view selection application 
%and semantic caching of query results. 
%and our results show $1.7\times$ improvement 
%in view selection and $8.6\times$ improvement in semantic caching 
%using forecasted workloads compared to historical workloads. %highlighting the efficacy of \sys.

\revision{%It is worth highlighting 
%We emphasize that 
The primary contribution of \sys is not proposing new ML algorithms, rather leveraging them for the problem of forecasting \textit{entire} query statements, adapting feature selection and combining encoder-decoder architecture with LSTM for better accuracy, and improving scalability through template cutting and packing. To the best of our knowledge, \sys\ is the first framework that learns and forecasts the \emph{entire} statements of future recurrent queries in various time spans.}

\section{Observations and Motivation}\label{sec:motivation}
In this section, we present our observations from multiple real-world workloads, which serve as the motivation for the development of \sys. We provide a brief description of these workloads below along with detailed statistics (see Table~\ref{tab:4statistics}).

%The motivation of \sys stems from our observation of diverse real-world workloads. Below, we perform a quantitative and qualitative study on four real workloads to illustrate our observations. 
%We perform a quantitative and qualitative study on three real-world database workloads in the following perspectives: re-occurrence, evolving patterns, and predictability. We mainly use the WT workload as an example to illustrate our observations and insights. The similar conclusions apply to the other three workloads.
%\subsection{Workloads}
%In this paper, we investigate four real-world database workloads: 

%We briefly describe these workloads below. Table~\ref{tab:4statistics} shows detailed statistics about these workloads.

% \tar{In our team, we typically do not give the actual names of the workloads, we just say Customer 1,  Customer 2, etc. For SCOPE, it's ok. Please double check.} \rana{ double-checked with team, we are good}

\vspace{1mm}
\noindent\textbf{\wmc: } This workload contains 14 days of point lookup queries from a decision support system used for querying telemetry data of %Windows 
\microsoft's products and services. %~\cite{wtcituscon}. 
%Most of queries compute some aggregate statistics on product telemetry filtered by various conditions. %version, OS build, etc. %The workload consists of queries from the decision support system for timely decisions and actions to continuously improve Windows products and services. Aggregate stats on device telemetry filter by time, version, build
%\jesus{TODO}\yuanyuan{@Jesus, please confirm the description.}

\vspace{1mm}
\noindent\textbf{\scope:} The workload contains 2 weeks of production jobs executed in \microsoft's \scope query engine. %~\cite{chaiken2008scope}. % on Cosmos~\cite{cosmos}, 
% which is an internal cloud data service in \microsoft. 
The jobs are %mostly big data analytical queries 
written in the \scope query language, some of which contain UDFs (user defined functions) and UDOs (user defined objects).

\vspace{1mm}
\noindent\textbf{BusTracker:} The workload contains 57 days of queries from a mobile phone application for live-tracking of the public transit bus system, open-sourced by~\cite{ma2018query}. %The queries help users find nearby bus stops and get route information. %The workload traces are open-sourced by~\cite{ma2018query}.
%The dataset is open-sourced by~\cite{ma2018query}.

\vspace{1mm}
\noindent\textbf{\mssales:} %\mssales\ 
The workload contains 32 days of analytical query traces from \microsoft's internal revenue reporting platform. %that manages revenue data related to \microsoft products and services. 
It consists of queries on purchase, sales, budget, and forecast data.%, some of which drills down for a more comprehensive view of transactional details.

\begin{table}[ht!]
  \caption{The basic statistics of the workloads %\rana{ Number of inputs is the number of tables?}
  }

    %\vspace{-5pt}
  \label{tab:4statistics}
    \resizebox{\linewidth}{!}{
  \begin{tabular}{c|c|c|c|c}
    \toprule
    & \textbf{\wmc} & \textbf{\scope} & \textbf{BusTracker} & \textbf{\mssales}\\
    \midrule
    %DBMS type & PostgreSQL & ? & PostgreSQL \\
    trace length (days) & 14& 14 & 57& 32\\
    \# queries & 2.6M & 6M & 25M& 13.3K\\
  \bottomrule
\end{tabular}
}
%   \vspace{-5pt}
\end{table}

\vspace{-3mm}
{
%\tiny
\begin{figure}[h!]
\centering
\begin{sqllisting}
(*@\textbf{\textit{Parameterized Query}}@*):
SELECT A.x,A.y, SUM(A.e) / SUM(A.z) AS val
FROM A
WHERE A.val1 = $1 AND A.val2 = $2 AND A.val3 IN $3
GROUP BY A.x, A.y
(*@\textbf{\textit{Parameters}}@*): $1='v1', $2='v2', $3=(1,2)
\end{sqllisting}

\caption{An example of parameterized query.}%\hanxian{Since we removed the `parameterized query' in \S3, shall we remove this figure or move it to \S5.1?} \rana{Moved to \S5.1}}
\label{lst:template}
\end{figure}
}

\viewpoint{Observation 1}{Queries in real workloads are highly recurrent.}

Observing real production workloads, we found that most queries come from applications that use programmatic parameterized queries, with an example shown in Figure~\ref{lst:template}. Many queries in the workload share the same query template, while the parameter values vary. \revision{%Following similar definition in prior 
Building on definitions from prior
work~\cite{jindal2021production,alibaba,ma2018query,abebe2022tiresias,CrocodileDB,p-store}, we call a query that shares the 
same template with at least another query as a 
\textit{recurrent query}.}
%We call these queries  \textit{recurrent queries}.
%In these queries, the query templates are shared have a separate query template, or query logic, from the parameter values used in the query. Many queries in a workload share the same query template but with different parameter values, making them recurrent queries. 
As depicted in Table~\ref{tab:recurrence}, we observe over $94.5\%$ of queries in the four workloads are recurrent. 
%\yuanyuan{Would it be possible to provide some stats to show that most reccurent templates have more than 2 queries? This is optional.}
 The templates of the recurrent queries are \textit{recurrent templates}. Table~\ref{tab:recurrence} also provides the number of recurrent templates in the four workloads. The BusTracker workload %, which serves a single application, 
 has a small number of recurrent query templates. In contrast, \scope %, being a general big data platform used by multiple applications within \microsoft, 
exhibits a large number of recurrent templates~\cite{jindal2018selecting}. %\tar{maybe cite prior work on SCOPE that made similar observations.}\hanxian{@Jesus, @Rana, could you help on this?}
\revision{We also observe the dominance of frequent recurrent queries, with more than 94\% of total queries in the workloads having more than 20 recurrences.}

\begin{table}[t!]
  \caption{Recurrent queries in the workloads%\rana{ Number of inputs is the number of tables?}
  }
%       \vspace{-2pt}
  \label{tab:recurrence}
    \resizebox{\linewidth}{!}{
  \begin{tabular}{c|c|c|c|c}
    \toprule
    & \textbf{\wmc} & \textbf{\scope} & \textbf{BusTracker} & \textbf{\mssales}\\
    \midrule
    %These are the updated data with definetion of recurrent "appear more than once"
    \% recurrent queries &99.9\%  & 94.5\% & 99.9\%  &96.9\% \\
    %Number of inputs &12 & 9733 & ?\\
    \# recurrent templates & 2157& 168197& 258 & 1143\\
  \bottomrule
\end{tabular}
}  \vspace{-1mm}
\end{table}

\viewpoint{Observation 2}{Recurrent queries often evolve over time.}

%\subsection{Evolving Patterns}\label{sec:evolvingPatterns}
%\tar{is it possible to classify/categorize the evolving patterns? I think you are already partially doing this in Figure 2, just state them here too.} \yuanyuan{Like what Tarique suggested, we need to define what "evolve" means in our context. Perhaps, here it means that query parameters change with query arrival time.}
% Real workloads are evolving rather than static over time. Even though many queries share the same query template, the parameter values often change with the query arrival time. For a recurrent template in a workload, if the parameter values for the queries belonging to this template change over time, we call the template as an \textit{evolving template}. Note that there can be multiple parameters in a query template, as long as one of the parameters exhibits changing behavior, the template is an evolving template. Queries in an evolving template are \textit{evolving queries}. Table~\ref{tab:evolution} shows the percentage of evolving templates among the recurrent templates, as well as the proportion of the evolving queries among all the recurrent queries in the four workloads. The majority of the recurrent queries in \wmc, \scope, and BusTracker are evolving. Although for \mssales the evolving queries only constitute $25.3\%$ of the recurrent queries, these queries take roughly $57.4\%$ of the total execution time of the entire workload. In other words, the evolving queries are the more expensive ones in the workload. 

In real workloads, the parameter values in recurrent templates can change dynamically over time.
%Even though many queries may share the same query template, the parameter values often change across queries. 
A recurrent template with at least one changing parameter value is called an \textit{evolving template}. 
%The template is considered an evolving template even if only one parameter exhibits changing behavior. 
Query instances belonging to evolving templates are referred to as \textit{evolving queries}. Table~\ref{tab:evolution} shows the percentage of evolving templates among the recurrent templates, %as well as 
and the proportion of evolving queries among the recurrent queries in the four workloads. In \wmc, \scope, and BusTracker, the majority of recurrent queries are evolving. In \mssales, evolving queries represent $26.6\%$ of recurrent queries but account for $57.4\%$ of the total workload execution time, %making them the most expensive queries in the workload.
indicating their higher cost.

\begin{table}[t!]
  \caption{Time-evolving queries in the workloads %\rana{ Number of inputs is the number of tables?}
  }
%       \vspace{-2pt}
  \label{tab:evolution}
    \resizebox{\linewidth}{!}{
  \begin{tabular}{c|c|c|c|c}
    \toprule
    & \textbf{\wmc} & \textbf{\scope} & \textbf{BusTracker} & \textbf{\mssales}\\
    \midrule
    \% evolving templates & 96.6\%& 97.3\%& 99.8\%& 0.2\%\\
    \% evolving queries & 99.9\%& 99.4\%& 99.9\%& 26.6\%\\
  \bottomrule
\end{tabular}
}
 %  \vspace{-5pt}
\end{table}

\begin{figure}[ht!]
  \centering
  \includegraphics[width=\linewidth]{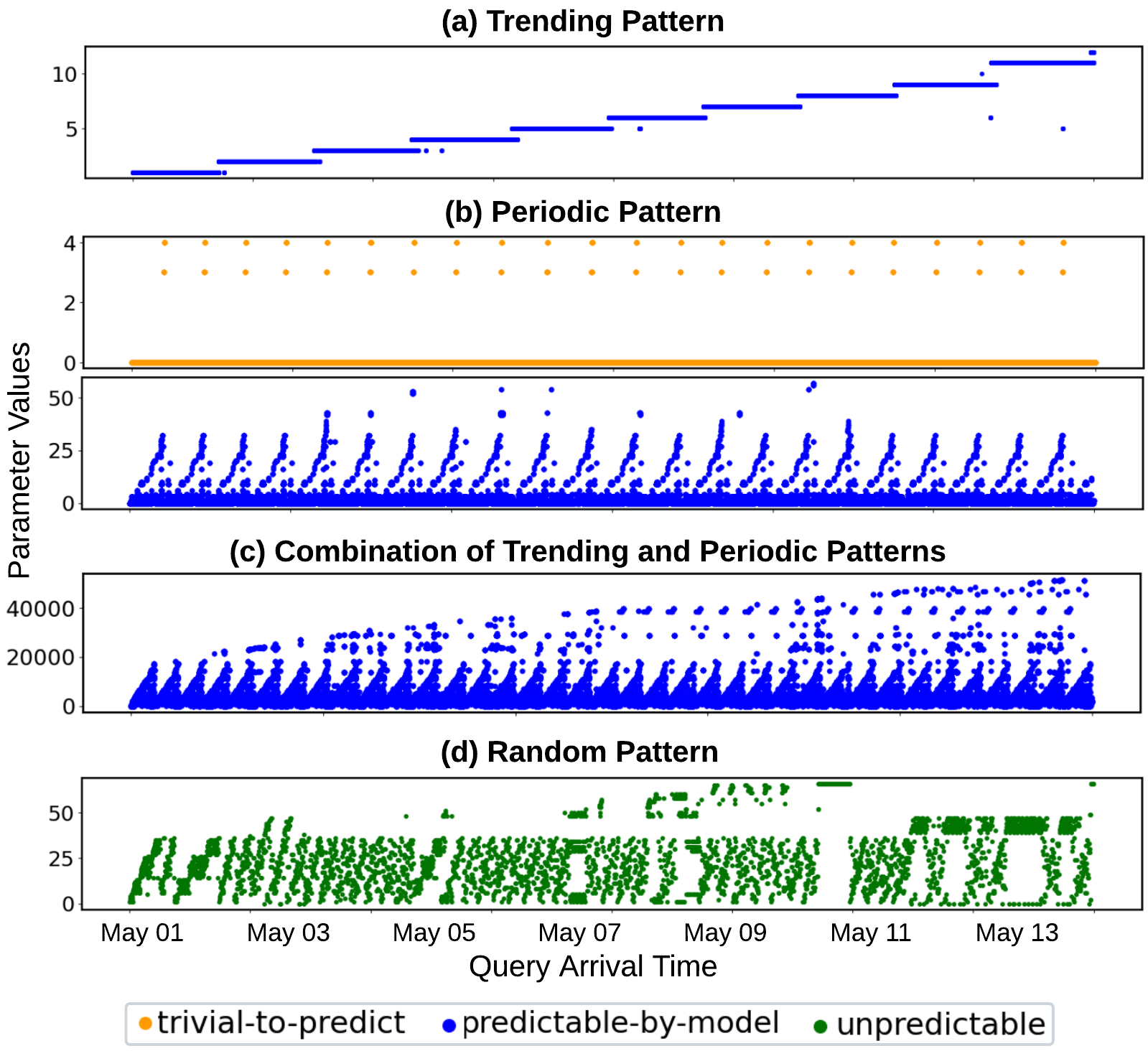}
  \caption{Characterizing time-evolving patterns and their predictability using queries from the \wmc workload. %\tar{unclear what the red rectangles are depicting. do we really need them?} %\yuanyuan{@Hanxian, can you rename the subfigures based on the new names of the patterns? Also, can we redraw the figures with arrival time as the x-axis instead of time sequence? Also remove the name of the parameters.}
%\rana{ To save space, add the legend at the top (horizontal legend), and consider increasing the size of the font}
}
  \label{fig:CommonPatterns}
 \vspace{-1mm}
\end{figure}

We further analyze parameter value changes with query arrival time and identify common patterns: (a) trending pattern: increasing, decreasing, or level trends; (b) periodic pattern: regular pattern with fixed interval (\eg hourly, daily, weekly); (c) combination of trending and periodic patterns; (d) random pattern (no regular or predictable pattern). Examples of these patterns are visualized in Figure~\ref{fig:CommonPatterns} using the \wmc\ workload. 

 \vspace{-1mm}

\viewpoint{Observation 3}{Recurrent queries are highly predictable.}

\begin{figure}[t!]
  \centering
  \includegraphics[width=\linewidth]{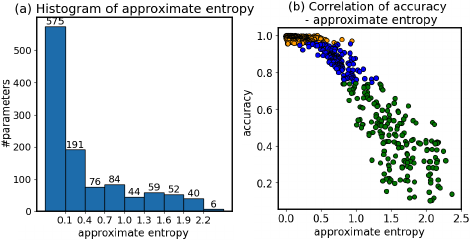}
  \caption{ (a) The histogram of ApEn on parameters of \wmc workload. (b) The negative correlation between parameter forecasting accuracy (using vanilla LSTM) and ApEn.
}
  \label{fig:ApEn_Pred}
\vspace{-2mm}
\end{figure}

We further study the predictability of parameters in the recurrent queries, by employing the widely-used approximate entropy (ApEn)~\cite{pincus1991approximate} metric to quantify the unpredictability of time-series parameter data. A lower ApEn indicates higher predictability. Figure~\ref{fig:ApEn_Pred}(a) shows the histogram of ApEn of all parameters of the \wmc workload. We observe some parameters have low ApEns, even smaller than $5e^{-4}$ in our study, while some have relatively higher ApEns, %which are 
greater than $1.0$. 

%\vspace{-2mm}

Unfortunately, it is well known that there is no universally defined threshold value for classifying a variable as unpredictable using ApEn~\cite{li2016estimation,chuckravanen2014approximate}. %\yuanyuan{@Hanxian, can you add a reference here?} 
It all depends on the context and the threshold is determined empirically for a particular scenario. However, not surprisingly, we found that ApEn is negatively correlated with parameter prediction accuracy by an ML model. Figure~\ref{fig:ApEn_Pred}(b) illustrates this relationship for a widely-used vanilla LSTM model~\cite{hochreiter1997long}. Empirically, we tried various ML models\footnote{We tried Random Forest, vanilla LSTM, and our own \sysModel.} for predicting each parameter, and get the best accuracy value $Acc_{max}$ for each parameter. To quantitatively identify unpredictable parameters, we introduce an accuracy threshold $\tau$, and treat parameters with $Acc_{max}$ lower than $\tau$ as unpredictable parameters. We empirically set $\tau=75\%$ based on our expectation on parameter accuracy. Among the predictable parameters, we further identify the trivial-to-predict ones, which have simple patterns, \eg repeating only a very small number of possible values. These can be easily predicted by simple heuristic based methods. The remaining predictable parameters are evolving over time with more complex patterns and require sophisticated models to achieve good forecasting accuracy. We use orange, blue, and green colors to mark the three categories: \textcolor{orange}{\textbf{trivial-to-predict}}, \textcolor{blue}{\textbf{predictable-by-model}}, and \textcolor{green!40!black}{\textbf{unpredictable}} in Figure~\ref{fig:CommonPatterns}.
Table~\ref{tab:Predictability} shows the percentages of these categories for the four workloads. Overall, our analysis indicates the high predictability of parameters in recurrent templates, with a significant portion of them non-trivial to predict, thus necessitating ML models for assistance.
\begin{table}[t!]
  \caption{The predictability of workloads}
      % \vspace{-1pt}
  \label{tab:Predictability}
  \resizebox{\linewidth}{!}{
    \begin{threeparttable}
  \begin{tabular}{c|c|c|c|c}
    \hline
    & \textbf{\wmc} & \textbf{\scope} & \textbf{BusTracker} &\textbf{\mssales}\\
    \hline
    trivial-to-predict & 21.3\% & 23.1\% & 13.7\% &9.1\%\\
    predictable-by-model & 68.5\% & 62.0\% & 78.7\% &81.8\%\\
    unpredictable & 10.3\% & 14.9\% & 7.6\% &9.1\%\\
    \hline
    \hline
    total predictable\tnote{$\dag$} & 89.7\%& 85.1\%& 92.4\%&90.9\%\\
    \hline
    
\end{tabular}
\begin{tablenotes}
    \footnotesize
    \item[$\dag$]Sum of the trivial-to-predict and the predictable-by-model categories
  \end{tablenotes}

\end{threeparttable}

}
% \vspace{-2pt}
\end{table}

Note that the four studied workloads represent a diverse set of workloads from different database systems with distinct query volumes (ranging from 13K to 25M) and varying numbers of query templates (ranging from 258 to 168K), as well as a mixture of operational queries (\wmc and BusTracker) and analytical queries (\scope and \mssales). Yet, the aforementioned observations remain true for all four workloads. In addition, these observations align with existing works~\cite{jindal2021production,alibaba,ma2018query,abebe2022tiresias,CrocodileDB,p-store} that emphasize the recurrent and predictable nature of real workloads. In fact, \sys's target applications, the various existing workload optimization tools, already implicitly assume workload predictability, as it simply doesn't make sense to tune performance on random workloads. Therefore, \sys is designed on the same premise of this common scenario.

%For example, both \cite{jindal2021production} and \cite{alibaba} observed the overwhelming evidence of recurrent queries and designed their materialized view recommendation methods based on this observation. The forecasting methods proposed in~\cite{ma2018query,abebe2022tiresias,CrocodileDB,p-store} all rely on their empirical observations that workloads are recurrent and predictable. Our \sys works on the same premise of these existing works, which present a common scenario as demonstrated by our study of four diverse workloads and the shared observation of existing works. In fact, the target application of \sys is the existing body of workload optimization tools, which already implicitly assume predictability of workloads (it simply doesn't make sense to tune performance on random workloads). 
% \begin{figure}[t!]
% \centering
% \begin{sqllisting}
% (*@\textbf{Query Q}@*):
% SELECT A.x, A.y, SUM(A.e) / SUM(A.z) AS val
% FROM A
% WHERE A.val1 = 'v1' AND A.val2 = 'v2' AND A.val3 IN (1,2)
% GROUP BY A.x, A.y

% (*@\textbf{\textit{Template temp(Q)}}@*):
% SELECT A.x,A.y, SUM(A.e) / SUM(A.z) AS val
% FROM A
% WHERE A.val1 = $1 AND A.val2 = $2 AND A.val3 IN $3
% GROUP BY A.x, A.y

% (*@\textbf{\textit{Parameters para(Q)}}@*):
% ('v1', 'v2', (1,2))
% \end{sqllisting}

% \caption{Templatization Example\hanxian{Since we removed the `parameterized query' in \S3, shall we remove this figure or move it to \S5.1?} \rana{Moved to \S5.1}}
% \label{lst:template}
% \vspace{-15pt}
% \end{figure}
%\section{Preliminaries and Overview} \label{sec:Problem}
%\vspace{-3mm}
\section{Problem Statement}
\label{sec:Problem} %rana{lock}

We begin by introducing  key concepts to  %that help us 
formally define the workload forecasting problem. 
% Before formulating the formal definition of the workload forecasting problem that we are solving, we first start with the definition of some basic concepts.
% We refer a \textit{query} $q$ as a statement written in SQL or a similar language, such as the \scope query language. It should be noted that the definition of a query in this paper is broad and encompasses insert, delete, or update statements as well, although the four workloads we study in this paper primarily contain query statements only.
A \textit{query}, denoted by $q$, refers to a statement expressed in SQL or a similar declarative language. %It is worth noting that our definition of a query is broad, encompassing not only \texttt{SELECT} statements, but also \texttt{INSERT}, \texttt{DELETE}, or \texttt{UPDATE} statements. However, this work primarily focuses on workloads that predominantly consist of \texttt{SELECT} statements. 
A \textit{workload}, denoted by $W$, is a bag (i.e. multi-set) of queries. Most workload-based optimizations tools~\cite{autoadmin,db2advisor} take $W$ as input, along with other constraints (e.g. storage constraint), and produce a recommendation of target features (e.g. indexes, materialized views, partitioning schemes, or MDC strategies) to optimize the total cost of $W$. These tools are usually executed at regular intervals, such as hourly, daily, or weekly. Hence, forecasting a representative query workload for the upcoming time interval necessitates %taking into account 
considering the arrival time of queries. As a result, we define a \textit{timed-workload}, denoted as $TW$, as a time series of queries, where each query $q_i$ in the workload has an associated arrival time $t_{q_i}$ indicating when the query is issued. Queries in the timed-workload are ordered based on their arrival time, i.e., $t_{q_1} \leq t_{q_2} \leq ... \leq t_{q_n}$. Hence, the timed-workload can be represented as $TW=[(q_1, t_{q_1}), (q_2, t_{q_2}), ..., (q_n, t_{q_n})]$. Given a timed-workload $TW$, its corresponding workload $W$ can be obtained by removing the arrival times and converting the sequence into a bag. When there is no ambiguity, we also refer to a timed-workload as a workload for simplicity.

We now present the formal definitions of two workload forecasting problems: next-$k$ forecasting and next-$\Delta t$ forecasting.

\begin{definition} (Next-$k$ Forecasting) \\ Given a timed-workload $TW=[(q_1, t_{q_1}), (q_2, t_{q_2}), ..., (q_n, t_{q_n})]$, the next-$k$ forecasting problem predicts the next $k$ future queries as:
$$F_k(TW) = [(q_{n+1}, t_{q_{n+1}}), (q_{n+2}, t_{q_{n+2}}), ..., (q_{n+k}, t_{q_{n+k}})]$$
\end{definition}

\begin{definition} (Next-$\Delta t$ Forecasting) \\ Given a timed-workload $TW=[(q_1, t_{q_1}), (q_2, t_{q_2}), ..., (q_n, t_{q_n})]$, the next-$\Delta t$ forecasting problem predicts the queries in the next time interval of size $\Delta t$ as:
$$F_{\Delta t}(TW) = [(q_{n+1}, t_{q_{n+1}}), (q_{n+2}, t_{q_{n+2}}), ..., (q_{n+\sigma}, t_{q_{n+\sigma}})]$$ where $t_{q_{n+\sigma}}<t_{q_{n}}+\Delta t \leq t_{q_{n+\sigma+1}}$  .
\end{definition} 

\begin{figure*}[t!]
%  \vspace{-2mm}
  \centering
  \includegraphics[width=\textwidth]{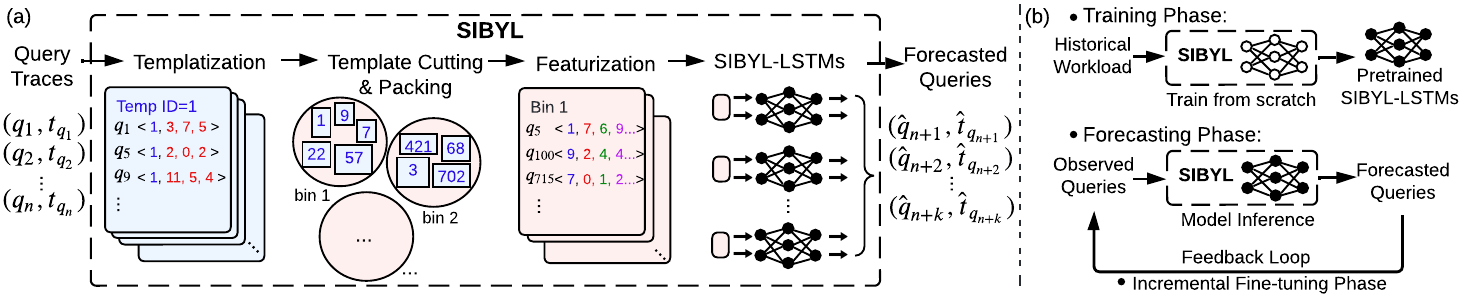}
  \caption{\sys\ Overview. (a) shows the four components of \sys for next-$\Delta t$ forecasting. 
  (b) shows the three phases of \sys. 
  %\yuanyuan{@Hanxian, can you remove the words online and offline from (b), as the word "online" can give people the impression that prediction is latency sensitive.}
  }
  \label{fig:overview}
%  \vspace{-5mm}
\end{figure*}

In contrast to the fixed number of queries to predict in next-$k$ forecasting, the next-$\Delta t$ forecasting deals with the prediction of queries within a fixed time interval. 
In real-world applications, determining the target time interval of the expected future queries is often more feasible and useful than the target number of future queries. 
Database tuning tasks are typically performed at regular intervals, such as hourly, daily, or weekly. Thus, having knowledge of the expected workload in the target time interval is crucial for optimizing database performance. 
%We define the next-$\Delta t$ forecasting problem as follows:

The next-$\Delta t$ forecasting problem is more challenging than the next-$k$ forecasting problem as the number of queries to predict, $\sigma$, is not known in advance and must be determined based on the arrival time of the predicted queries. 
Operationally, the forecasting model needs to keep predicting queries until it sees a query with an arrival time exceeding the next $\Delta t$ time interval.  

It is important to note that even though many workload optimization tools do not require time information, both forecasting problems defined above generate timed workloads. The predicted arrival time of each query plays a crucial role in forecasting, as it provides constraints for the next-$k$ (selecting the next k queries ordered by the arrival time) and next-$\Delta t$ (producing queries with bounded arrival time) prediction challenges. After the forecasting process, the timed workloads can be easily converted into regular workloads before being utilized by workload optimization tools.

\revision{We want to emphasize that this work
primarily focuses on analytic workloads targeting workload-based optimization applications. While our techniques are general enough, we leave forecasting non-query statements (DML and DDL) as future work.}

\section{\sys\ Overview} \label{sec:overview}

%\sys is a general-purpose workload forecasting framework that can be applied for a variety of use cases. It takes past query traces and predicts future query traces. To address this workload forecasting problem, we utilize machine learning models to perform time series prediction. The forecasted future queries with their arrival time can then be directly fed into any existing database administrative tools that expect a workload as input for various database optimization and tuning tasks. %To accommodate specific applications, \sys offers the flexibility to \textit{customize} accuracy measurement and objective functions of ML models that can be plugged into the system. 
\sys is an ML-based workload forecasting framework that takes the past queries as input and outputs the future queries, performing time series prediction. %These 
Forecasted queries can be seamlessly integrated into existing database tools for %efficient optimization and 
tuning tasks. 

We first tackle the simpler next-$k$ forecasting problem, by developing a template-based featurization method and combining stacked LSTM with an encoder-decoder architecture to create \sysModel. To address practical challenges such as handling a large number of queries in the required prediction interval and the scalability of the design, we extend the solution for the next-$k$ problem to the next-$\Delta t$ forecasting problem, and implement template cutting and packing algorithms to reorganize templates into bins. This allows us to build per-bin models, resulting in a more practical and scalable design. Figure~\ref{fig:overview}(a) depicts the overall architecture of \sys for solving the next-$\Delta t$ problem, and the detailed design of each component %of \sys\ 
will be introduced in \S\ref{sec:design-k} and \S\ref{sec:design-delta}.
%\hanxian{Until now, we have introduced all the components of SIBYL, maybe we can add a few lines or a subsection to summarize them and reference fig4(a):}
%\hanxian{ With template \cutNpack and adapting per-template model to per-bin model, \sys\ is now able to solve the next-$\Delta t$ problem consisting of four components as shown in Figure~\ref{fig:overview}(a). Given a workload, \sys\ first employs templatizations (\S\ref{sec:templatization}) for all the queries, and estimate the template sizes of target time interval $\Delta t$ for all templates (\S\ref{sec:tempalteSize}). \sys\ then performs template \cutNpack into bins (\S\ref{sec:cut}, \ref{sec:pack})and performs featurization for each bin (\S\ref{sec:featureEngineer}, \ref{sec:perbinmodel}). It finally trains and employs \sysModel\ (\S\ref{sec:templatemodels}) to forecast future queries in the next-$\Delta t$ for per bin. We combine the forecasted queries in all bins and sort them by the forecasted arrival time to get the next-$\Delta t$ forecasting for the whole workload.}

As shown in Figure~\ref{fig:overview}(b), \sys has the following three phases: 

\noindent
\textbf{Training%phase
:} %\sys 
it featurizes the past queries and their arrival time, and trains ML models from scratch. The training is only performed once. We assume that the historical workload provides sufficient training data for accurate predictions using ML models.
%Sufficient training data is needed for accurate predictions when using ML models. In this paper, we assume that the historical workload provides enough training samples. 

\noindent
\textbf{Forecasting%phase
:} %\sys 
it continuously receives recent queries from the workload traces and employs the pre-trained ML models to predict the next-$k$ or next-$\Delta t$ queries with their expected arrival time in the future workload. The forecasted workload can be passed to database tuning tools as input to perform optimization.
    
\noindent
\textbf{Incremental fine-tuning%phase
:} %\sys\ also provides a feedback loop to monitor the forecasting accuracy and detect the workload shifts (\eg new types of queries emerging in the workload). Based on the feedback, \sys\ automatically and efficiently adjusts the models to deal with workload shifts by incrementally fine-tuning the models on the shifted workload, without the need to train the models from scratch. %\tar{we should highlight this as an additional difference wherever we compare with related work.}
%\sys\ 
it monitors model accuracy and detects %shifts in 
workload shifts (\eg new types of queries emerging in the workload) via a feedback loop. It adjusts its models efficiently by fine-tuning incrementally on the shifted workload, without retraining from scratch.

\revision{Note that the three phases are not on the critical path of workload-based optimization applications. They are offline steps to prepare the inputs for these applications to tune database performance.}

\section{Next-$k$ Forecasting Models}\label{sec:design-k}
%\tar{give a brief overview of what you are going to tell in this section.}

We first solve the next-$k$ forecasting problem. A simplistic approach would be constructing a single global model to learn on all the queries in a workload. However, this method has various limitations. Firstly, as demonstrated in Table~\ref{tab:recurrence}, actual workloads frequently consist of different types of queries (different query templates), making it challenging to featurize the extensive range of query logic. Secondly, it results in a large number of features and a complex mixture of patterns from all templates that the model needs to learn. Thirdly, such a global model necessitates a substantial amount of training data and can be extremely expensive to train. 
%Previous research has the similar observations and discussion that the global model is not the perfect choice~\todox{cite}. \hanxian{@Rana}

Based on the observations outlined in \S\ref{sec:motivation}, %we understand that 
real-workload queries are highly templatized, and the query parameters in each template frequently exhibit time-evolving patterns and are very predictable. Rather than creating a global model for the entire workload, it is more reasonable to build a model for each template. Specifically, we perform query templatization (\S\ref{sec:templatization}) and group the queries in a workload based on their templates. For each template, we collect queries of the template, featurize the queries (\S\ref{sec:featureEngineer}), and train a model (\S\ref{sec:templatemodels}) to forecast future $k$ queries for this template and their arrival time. This high-level idea of constructing a model per template is also utilized in existing works such as \cite{ma2018query, cardlearner, costmodel}.
Note that $k$ is the forecasting window size, indicating how many queries are expected to forecast. We defer the discussion on practical considerations in selecting $k$ to \S\ref{sec:design-delta}. To get the forecasting results for the entire workload, we collect the forecasted queries for all templates together and sort them by the forecasted arrival time to produce the final $k$ queries. We next elaborate the design details.%As we will demonstrate below, this \textit{per-template} approach not only significantly simplifies the feature engineering, model training, model prediction, and model fine-tuning processes, but it also provides better prediction accuracy, as we show in \S\ref{sec:eval}. \tar{please double-check we have a baseline for global model?}\hanxian{we did not compared with the global model in eval. We tried a global model for WMC in summer (the per-table model) and had a 77.3\% recall. Not sure if we have space to compare and discuss it.} 
%\rana{We can cite SCOPE cardinality estimation paper, which discusses the issues with the global models}.

%\vspace{-2mm}
\subsection{Query Templatization}\label{sec:templatization}
In this pre-processing step, we group the queries in the given workload based on their query templates. 
%Although it is possible to rely on the application layer to tag queries based on the template they were generated from, our objective is to propose an approach that is application-agnostic and can work seamlessly with existing applications. 
%Nevertheless, \sys can accommodate other techniques for grouping queries based on their templates.
To obtain the template from a given query, \sys parses the query to create an abstract syntax tree (AST) and transforms the literals in the AST into parameter markers~\cite{parametermarkers}\footnote{We note that the extraction of templates does not take into account any query rewriting or optimizations.}. Our templatization technique can extract literals in any part of the query, including UDFs, UDOs, and Common Table Expressions (CTEs).
%\tar{are the literals restricted to WHERE and WITH clause, one may wonder if we can do this over the textual representation without parsing into ASTs?}.\hanxian{@Jesus} 
To ascertain the equivalence of two query templates, we verify whether their ASTs are identical via strict matching. 
Then, \sys associates all parameter bindings in a query to its corresponding template.
%Then, our approach associates all parameter bindings in queries that correspond to the same query template with that specific template. 
%Figure~\ref{lst:template} shows an example of templatization. 
%\tar{two equivalent queries can have different ASTs, maybe we should clarify that we are performing strict matching.}\hanxian{@Jesus}
%It is worth noting that 
%Our current implementation only supports strict matching to recognize equivalence. 
We rely on the AST representation of queries for templatization, as it simplifies the manipulation of the query structures. But our approach for evaluating template equivalence could be easily extended, \eg using canonical representations of filter expressions.
%Our approach for evaluating query template equivalence could be easily extended, particularly since we rely on the AST representation of queries, which simplifies the manipulation of their structure. 
%For example, we could use canonical representations of filter expressions. 
We leave such extensions for future work. 

%\vspace{-2mm}
\subsection{Feature Engineering}\label{sec:featureEngineer}
%\tar{some  major restructuring below.}\hanxian{looks good to me.}

%Feature engineering is crucial in forecasting future workload, and the challenge is to determine how to featurize queries and their arrival time. 
Featurization of queries with arbitrary complexity is a hard problem. Existing plan-based~\cite{marcus2019neo,marcus2022bao} or token-based~\cite{jain2018query2vec} featurization methods have made significant progress on this front, but can still only support limited query constructs. However, in our setting, by leveraging  query templates, we can \textit{circumvent} this challenge. Queries with the same template share identical query structure but differ only in their parameter values. Therefore, we can use the template identification (template id) to capture the query structure, while hiding the query complexity within the template. 

%It's worth noting that 
Each query can be featurized using template id and parameter values. This also simplifies the query reconstruction process by merely filling in the parameter values into the corresponding template to reconstruct a predicted query. 
%The template-based featurization method has also been adopted by QueryBot 5000~\cite{ma2018query} and some other ML-for-DB works~\cite{cardlearner, costmodel}. 
In addition, there are two requirements for featurization that we need to satisfy in our setting.
\squishlist
\item 
\textbf{R1}: To better learn the time-evolving patterns, time-related features, either the query arrival time or any query parameter value of Date-Time type, need to be encoded in a way to capture the periodicity (such as year, month, week, day, etc.) and seasonality (including weekdays, weekends, holidays, etc.) of time. 
\item 
\textbf{R2}: Since future observations in a time series problem are often dependent on past observations, featurization should capture the relationship between consecutive queries. 
\squishend
Below, we describe how we featurize a query, including its arrival time to meet these requirements.

\vspace{-1mm}
\subsubsection{\textbf{Query Feature Vector}.}
\label{sec:queryFeatureVector}
In a per-template model, only the parameter values require encoding in the query feature vector. Furthermore, to satisfy R2, we also encode the difference between the parameter values of the current query and its predecessor. Here the major task is on dealing with a rich set of data types for query parameters. We utilize the table schema and parameter values to deduce the data types of the parameters. Subsequently, we encode each parameter according to its corresponding data type.

\squishlist
\item \textbf{Numerical types}. The parameters of numerical-type, \eg \textit{Int}, \textit{Long}, \textit{Double} and \textit{Float}, are encoded by their numerical values.
\item \textbf{Categorical types}. The parameters with \textit{String}, \textit{Char}, \textit{Boolean} types are encoded as categorical values. For each parameter, we collect all possible values from the training data, and assign an identical integer value to each category. \revision{To deal with high cardinality categorical features, we then apply the %widely used 
feature hashing technique~\cite{featurehashing} to encode the categorical values.} %\tar{maybe explain how we calculate difference for these types?}
\item \textbf{Date-Time type}. Special treatment is given to date-time parameters to ensure R1. The value of date-time is dissected into individual components such as year, month, day, hour, minute, and second. Furthermore, we incorporate additional derived features such as identification of weekends, public holidays, the season of the year, and so on.
\item \textbf{Set type}. Set parameters often appear in the IN or VALUES clauses. In our study, we observe a fixed set of values recurring which are extracted as categorical values. %We plan to explore other ways of encoding set parameters in the future.
We plan to explore alternative encoding methods in the future.
\squishend

\vspace{-1mm}
\subsubsection{\textbf{Arrival Time Feature Vector}.} Analogous to processing date-time parameters, we decompose the query arrival time into its constituent parts, including year, month, day, hour, minute, and second, thus satisfying R1. This approach allows the ML models to forecast all the features related to time and reconstruct future arrival timestamps. In addition, to ensure R2, we also featurize the difference between the arrival time of successive queries.

\begin{comment}
For query arrival timestamp forecasting, the timestamp feature map consists of a sequence of timestamp feature vectors, with each feature vector obtained from a past query arrival timestamp and consists of two parts:

\noindent(1) The arrival timestamps for the query. 
%Since we also want to predict the arrival timestamps of the future queries, we also include the arrival time information collected from query logs. 
Similarly, we use the time-related features: the day of year, the day of week, hour and minute of the arrival timestamp. This enables the ML models can predict all these time-related features to reconstruct the future arrival timestamps accordingly.

\noindent(2) The differences between the consecutive timestamps. It helps ML models to learn how the collected query timestamps are changing affected by the previous one.
\end{comment}

\vspace{-1mm}
\subsubsection{\textbf{Input Feature Map for ML Models}.} As we formulate the workload forecasting problem as a time series prediction problem, the input to each ML model is a feature map with a sequence of feature vectors until the current timestamp ($fv_1, fv_2, ..., fv_n$) ordered by query arrival time. Each feature vector $fv_i$ comprises the query feature vector concatenated with the corresponding query arrival time feature vector. The output of each ML model is a sequence of the next $k$ feature vectors ($fv_{n+1}, fv_{n+2}, ..., fv_{n+k}$), which are %then 
used to reconstruct the next $k$ queries with their respective arrival times. %\tar{the next line is not very clear because the LSTM model may be able to capture the interrelationships irrespective of the feature map? maybe given an example to illustrate what we mean.} 
The input feature map enables the ML models to capture the interrelationship among features within and between queries and learn from these connections, resulting in accurate predictions.

\subsection{Forecasting Models}\label{sec:templatemodels}

We now outline the ML models for solving the next-$k$ forecasting problem. With various options available for time series forecasting, we have considered and evaluated the following two models:
%\tar{we need to justify why the below few models and not something like Transformer? One option is to add a paragraph after LSTMs acknowledging that other models such as Transformer can also be used, but we limit to LSTMs since it gives sufficiently accurate results as we show in experiments.}

%\hanxian{shall we move the baselines to the evaluation section?} \yuanyuan{Yes.}

\vspace{1mm}
\noindent\textbf{Random Forest (RF).}
RF is an ensemble learning method~\cite{sagi2018ensemble} widely used for classification and regression problems. 
%It first splits the features by bootstrap sampling and construct one decision tree to deal with each subsample. Every decision tree makes its individual decision based on the subsample features and finally all the results from the decision trees are aggregated and averaged out to be the final result. 
With its simplicity and popularity, it is tempting to apply %this simple model 
RF to our forecasting problems. The conventional use of RF enables the prediction of only the next single query. To adapt RF for predicting next-$k$ queries $q_{n+1}, q_{n+2},...q_{n+k}$, we use the past $k$ queries $q_{n-k+1}, q_{n-k+2},...q_{n}$. More specifically, RF predicts $q_{n+1}$ using $q_{n-k+1}$, $q_{n+2}$ using $q_{n-k+2}$, and so on. 
%RF is a simple model that is easier to train. 
However, as we will show in \S\ref{sec:eval}, even with this adaption, RF is still not a good solution for our problem. 
%the unsatisfied results shown in \S\ref{sec:eval}. 

\begin{figure}[t!]
  \centering
  \includegraphics[width=\linewidth]{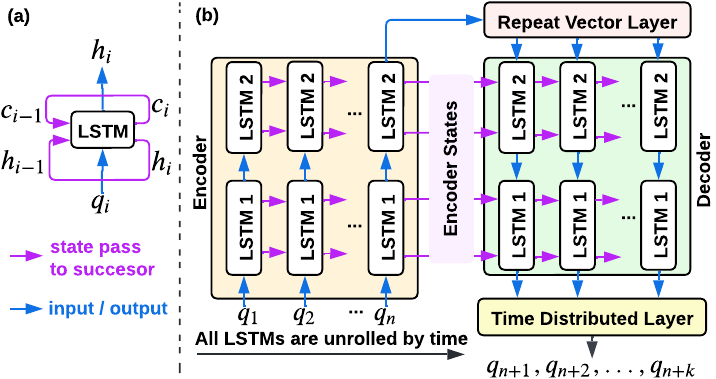}
  \caption{(a) One LSTM layer. (b) \sysModel. %\tar{Typo in Repeat Vector Layer}
  }
  \label{fig:e2d2}
% \vspace{-10pt}
\end{figure}

\vspace{1mm}
%\hanxian{I rewrited the next two paragraphs. Added the limitation of LSTM and the reason of our design.}
\noindent\textbf{Long Short-Term Memory Networks (LSTM).}
LSTM is a %more 
variant of the recurrent neural network (RNN)~\cite{medsker2001recurrent}, designed to learn a sequence of data with both short-term and long-term dependencies. Its ability to capture temporal patterns makes it %well-suited 
ideal for solving time-series prediction problems. However, vanilla LSTM model, \ie a single-layer LSTM as shown in Figure~\ref{fig:e2d2}(a), has limited model capacity to capture complex relationships among features. Our task involves taking a sequence of historical queries and forecasting a sequence of future queries, with each query consisting of multiple parameters, making it a \textit{multi-variate, multi-step, time-series sequence-to-sequence} learning problem. While LSTM is good at capturing temporal dependencies, it processes each variable independently and does not directly capture the inter-dependencies between them. Thus, it is not a good solution for multi-variable problem that involves complex relationships among variables. To address this limitation, we combine LSTM with the advanced encoder-decoder architecture~\cite{cho2014learning}, resulting in a model that we refer to as \sysModel, to better capture both the temporal dependencies and the interrelationships among variables, as shown in Figure~\ref{fig:e2d2}(b). %\tar{we should mention the last line as our novel contribution where we talk about \sysModel in the intro}.

The encoder-decoder architecture is widely used in %natural language processing
NLP domain ~\cite{cho2014learning,lu2016training} due to its ability to capture the context more effectively. As shown in Figure~\ref{fig:e2d2}(b), the encoder processes the input sequence and maps features into encoder states. The encoder states are the latent representations that summarize the entire input sequence encoding important information and dependencies from the input. The encoder states are then used to guide the generation of the output sequence by the decoder. This architecture allows the model to capture complex relationships between the input and output sequences and performs better on complex evolving patterns.

%\yuanyuan{@Hanxian, can you take a look at the following paragraph? I removed the argument on incremental training since there seems to be some work on it.}
%When comparing RF and LSTM, it is evident that RF is a simpler model that is easier to train. However, even with our adaption, it is fundamentally unsuitable for time series sequence-to-sequence prediction and has unsatisfied results in our problem. 
As will be shown in \S\ref{sec:eval}, \sysModel produces more accurate and stable results on different problem settings (\ie various $k$ and $\Delta t$ settings) than RF and vanilla LSTM. 
%Moreover, \sysModel allows incremental training and fine-tuning. 
Therefore, we select \sysModel as our preferred model. We %further 
note that the existing forecasting models in~\cite{ma2018query,meduri2021evaluation} are not directly suitable for our needs: QueryBot 5000~\cite{ma2018query} performs single-variable, multi-step forecasting, while \cite{meduri2021evaluation} conducts multi-variable, single-step forecasting.
%However, it has a significant drawback in that it does not support incremental training and fine-tuning\tar{double-check there is no work on this or its variants? otherwise I would say it has limited support .... compared to LSTM.}\hanxian{RF is a conventional ML method based on statistics, not a deep learning method, so RF and its variants are not supporting incremental learning.}. Conversely, LSTM is more computationally expensive to train but allows incremental training. Moreover, as demonstrated in \S\ref{sec:eval}, \sysModel exhibits higher accuracy in predicting workload, which led us to select it as the preferred model for solving the workload forecasting problem. 
%We further note that ML models used in previous workload forecasting papers~\cite{ma2018query,meduri2021evaluation} are less suitable for our setting where we perform multi-variable and multi-step time-series forecasting\tar{double check CMU paper is less suitable for multi-variable multi-step prediction. I guess they can do so? If we are not 100\% sure we should tone down}
%\hanxian{Is it better to say: "we did not directly compare with ~\cite{ma2018query,meduri2021evaluation} because the original usage of their model are for multi variable-single step forecasting~\cite{meduri2021evaluation} and single variable-multi step forecasting~\cite{ma2018query}, which are different from our settings." ?}. 
Finally, large language models~\cite{vaswani2017attention,devlin-etal-2019-bert,radford2019language} could also be considered. % While these models typically perform well on complex sequence-to-sequence learning problems, they come with a trade-off: they include more parameters and take longer to train. 
 These models excel in complex sequence-to-sequence learning but have more parameters and longer training time.
Based on our evaluation in \S\ref{sec:eval}, \sysModel provides sufficiently accurate results, so we chose to focus exclusively on \sysModel.

\subsubsection{\textbf{\sysModel}}
Figure~\ref{fig:e2d2} shows how we combine LSTM with an encoder-decoder architecture to build \sysModel.
%As mentioned above, we combine LSTM with an encoder-decoder architecture to build \sysModel for our setting, as shown in Figure~\ref{fig:e2d2}.
%The encoder-decoder architecture is suitable to solve the sequence-to-sequence learning problems, \ie the input and output are both sequences.
%, in Natural Language Processing (NLP), e.g., machine translation~\cite{cho2014learning}, speech-to-text recognition~\cite{lu2015study,lu2016training}. %Since our target is forecasting a window (multi-step) of future queries with each query has multiple parameters (multi-variate), we formulate our multi-variate multi-step time-series forecasting problem into a sequence-to-sequence learning problem. 
%The encoder processes the input sequence and maps all the features into the encoder states as shown in Figure~\ref{fig:e2d2}(b). The encoder states are the latent representation that captures the important information from the input. The encoder states are then fed into the decoder to generate the desired output. In this paper, we adapt the encoder-decoder architecture and the stacked LSTMs into \sysModel as shown in Figure~\ref{fig:e2d2}. %\hanxian{TODO: rewrite+explain, more intuitive}

We first briefly introduce how one LSTM layer works, as shown in Figure~\ref{fig:e2d2}(a). At each time step $i$, the LSTM takes an input $q_i$ and produces a hidden state $h_i$. Additionally, it passes two types of states to the next time step: the hidden state $h_i$, which governs short-term memory, and the cell state $c_i$, which governs long-term memory. The LSTM works in a recurrent manner, which can be viewed as multiple copies of the same network at different time steps, with each passing the information (the states) to a successor. Figure~\ref{fig:e2d2}(b) depicts the unrolling of the LSTM over multiple time steps, allowing the model to retain long-term and short-term information and enabling it to reason about prior data to inform subsequent ones.

Integrating LSTM with the encoder-decoder architecture, we design \sysModel in Figure~\ref{fig:e2d2}(b). We first deploy a stacked-LSTM with two LSTM layers as the encoder. The stacked LSTM is deeper and provides higher model capacity (more trainable parameters) to capture more information at different levels and model more complex representation in data, compared to the vanilla LSTM. The first layer of the encoder (\textit{LSTM 1}) takes featurized query $q_i$ at time step $i$, together with the previous cell state $c_{i-1}^1$ and hidden state $h_{i-1}^1$. The second layer (\textit{LSTM 2}) takes the output from \textit{LSTM 1} as input, along with its previous cell state $c_{i-1}^2$ and hidden state $h_{i-1}^2$. The encoder recursively learns on the entire input sequence and generates the output. The final encoder state, which summarizes the input sequence, consists of $h_n^1$, $c_n^1$, $h_n^2$, and $c_n^2$. We then employ a repeat vector layer to replicate the encoder output into $k$ copies. The decoder module comprises two LSTM layers that are initialized with the final encoder state. The decoder takes the encoder output together with the previous states as the input, to generate the hidden state output for each of the $k$ future steps. Finally, we apply the time distributed layer, which is a dense layer, to separate the results into each future time step. The output of \sysModel is a sequence of feature vectors with length $k$ which can be decoded into $k$ future query statements and their arrival time.
\section{Next-$\Delta t$ Forecasting Models}
\label{sec:design-delta}

%In the previous section, we discussed our approach to solving the next-$k$ forecasting problem. As illustrated in \S\ref{sec:Problem}, a more practical forecasting benefits DBMS applications is acquiring the future queries in the next time interval $\Delta t$. Now, 
We now address the practical challenges in adapting next-$k$ forecasting to next-$\Delta t$ forecasting, for a fixed target $\Delta t$ value.

%we now extend its solution to solve the more demanding next-$\Delta t$ forecasting problem. We first motivate the need for 

%\vspace{-2mm}
\subsection{Challenges}% in Adapting Next-$k$ Forecasting for Next-$\Delta t$ Forecasting}

\noindent\textbf{Challenge 1: A required forecasting size exceeds a feasible $k$.} For a target time interval $\Delta t$, a naive approach would be to use the next-$k$ forecasting method (\S\ref{sec:design-k}) and set a $k$ sufficiently large so that $k \geq \sigma$, where $\sigma$ is the number of queries arriving in next $\Delta t$ as defined in \textsc{Definition 3.2}. Specifically, we refer to the number of queries for template $temp_i$ in $\Delta t$ as its \textit{template size} in $\Delta t$, denoted as $\sigma_i(\Delta t)$. %Specifically, we can first predict the next $k$ queries $q_{n+1}, q_{n+2}, \dots, q_{n+k}$ for each template. We then find the first predicted query $q_{n+m+1}$ that exceeds the $\Delta t$ window based on the predicted query arrival time, i.e., $t_{q_{n+m}} < t_{q_{n}} + \Delta t \leq t_{q_{n+m+1}}$, where $m+1 \leq k$. Finally, we return $q_{n+1}, q_{n+2}, \dots, q_{n+m}$ as the output for the next-$\Delta t$ queries of the template. \hanxian{this part is mentioned in S3.}
%To obtain the forecasted queries for the entire workload, we sort and merge the forecasted queries of all templates according to their predicted arrival times. 
However, as we will discuss below, it is not always practically feasible to fulfill $k \geq \sigma_i(\Delta t), \forall temp_i$. %On one hand, $k$ cannot be arbitrary large and is fixed during model training and inference. On the other hand, a large chosen $\Delta t$ (\eg 1 day or 1 week) for a workload with high volume could result in a large number of queries to forecast.

%real workloads can have large volumes of queries as shown in Table~\ref{tab:4statistics}, while applications can choose a large $\Delta t$, \eg 1 day, 1 week, which results in a large required number of queries to forecast in the next $\Delta t$.

%\tar{add a para that explains that we keep k fixed, and why it cant be too big or too small. also how we determine it or point to the relevant places where we discuss it. then the next two challenges will become clear}
First, the output window size $k$ of a sequence-to-sequence learning model is usually decided empirically rather than arbitrarily chosen, considering several factors: (1) %The practical factor is the machine resource limitation. 
A larger $k$ brings more computation complexity and memory overhead. Given a specific machine to deploy \sysModel\ and a certain-sized workload, the maximum feasible $k$ is decided and bounded by the machine resources. (2) The model accuracy degrades %when 
with arbitrarily increasing $k$, as it is harder for a model to capture the patterns in an extremely long sequence. %Today's 
The state-of-the-art sequence-to-sequence learning models~\cite{devlin-etal-2019-bert,radford2019language} typically adopt a feasible $k \in [128, 1024]$. %Previous research~\todox{cite} attempt to enlarge $k$ to $\sim 4000$ with additional training tricks to maintain accuracy, but it is still not arbitrary large or not even able to cover the largest template with $10{,}500$ queries per day in WMC, as shown in Figure~\ref{fig:imbalanced}. 
(3) A larger $k$ requires more training data, thus may lead to the problem of inadequate data available for training. 
%This is particularly true for time-series data, where the length of the time series limits the number of possible forecasting windows. 
Therefore, a maximum feasible $k$ in \sys is decided by an exhaustive search of the available resources, workload sizes, and model accuracy expectation once for a given experimental set-up.
%Thus $k$ is required to carefully decided rather than an arbitrary large value according to the applications. 
Once $k$ is decided, it is fixed during the whole training and inference process. A change of $k$ leads to a new model and requires model retraining from scratch.

\begin{figure}[t!]
%  \vspace{-5pt}
   \centering
   \includegraphics[width=\linewidth]{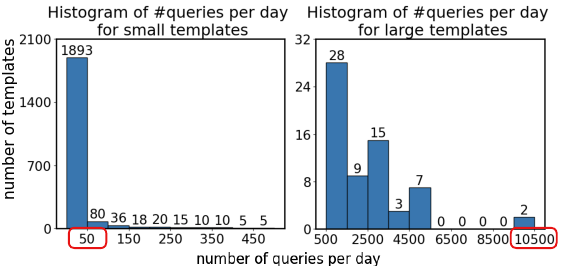}
   \caption{An example from the \wmc workload where the average number of queries  per day is imbalanced.} %\rana{ I don't think you need to rephrase the proposed approach here again the Figure since you are not showing the bins. Removing this will give some space.} To address this, we propose splitting  larger templates and grouping smaller templates to fit the model forecasting window.%\yuanyuan{This figure needs to be changed} 
% \vspace{-5pt}
   \label{fig:imbalanced}
 \end{figure}

%Now since we keep $k$ fixed, 
Secondly, given $k$ is limited and fixed, %there are two challenges when adopting the next-$k$ solution. First, 
the required number of queries to forecast for next-$\Delta t$ can be larger than $k$. For example, a physical design tool may require next one-day's queries to perform optimization, \ie $\Delta t= 1 day$.
%considering a target %time window of one day, i.e. 
%$\Delta t= 1 day$ (a common time interval for physical design tasks), 
As shown in Figure~\ref{fig:imbalanced}, the required forecasting window size for a large sized template in the \wmc workload can be up to $10{,}500$ to cover next day's queries. In this case, directly applying the next-$k$ solution is impossible with the above discussed feasible setting of $k \in [128, 1024]$.
%First, the query arrival rates of different templates can vary significantly. 
%As illustrated in Figure~\ref{fig:imbalanced}, the largest template in the \wmc workload has on average $\sim 10k$ queries per day, while the smallest one only has $\sim 10$ queries. 
%When the target time window is one day, i.e. $\Delta t= 1 day$ (a common time interval for physical design tasks), the per-template model for the largest template has to predict $\sim 10k$ queries, assuming a stable daily query volumes. 
%However, even when trained on massive clusters of GPUs, today's sequence-to-sequence learning neural networks can typically maintain only a $512$ to $2048$ forecasting window size~\cite{devlin-etal-2019-bert}. 
%So, for real workloads, it is not always feasible to increase $k$ to cover all the queries in the next $\Delta t$ window.  
This leads to the \textit{first challenge}: addressing \textit{large templates}, \ie templates with a substantially larger number of expected queries than $k$ in next $\Delta t$.

\vspace{1mm}
\noindent\textbf{Challenge 2: Per-template model solution is not efficient for small templates and not scalable.} Conversely, the \textit{second challenge} involves the long tail of \textit{small templates}, i.e., templates with a very limited number of expected queries in next $\Delta t$. For instance, Figure~\ref{fig:imbalanced} shows there are $1{,}893$ templates in the \wmc workload, which is $87.8\%$ of all the templates, with fewer than $50$ queries per day, much smaller than a common setting of $k$, \ie $k \gg \sigma_{small\_temp}(\Delta t)$. Directly deploying the next-$k$ solution is inefficient because it does not make the best use of model capacity. Moreover, in real workloads, the number of templates can be quite large, as shown in Table~\ref{tab:recurrence}. Training a model for each template is not scalable, resulting in a vast number of models to maintain, significant training time, and model storage overhead (see Table~\ref{tab:estimatedPerModel}).

%We next discuss how we address these two challenges.

%Given the resources available on the machines, we first estimate the maximum possible forecasting window size~\todox{{cite}} and determine a feasible value of $k$, both of which are workload-independent constants. The algorithm involves two main operations:

%\subsection{\cutNpack Algorithm}
%We introduce a two-step algorithm called \cutNpack, to address the above two challenges correspondingly. In the first step, the algorithm cuts larger templates into smaller templates of size smaller or equal to $k$. Then, in the second step, it packs multiple templates with size smaller than $k$ together into a bin (a logical term denoting a group of templates) such that the size of the bin is bounded by the bin capacity $k$. We discuss the two steps in more detail below.

%\vspace{-2mm}
\subsection{Template Cutting} \label{sec:cut}
We first address the problem with large templates with sizes greater than $k$ in Challenge 1. %\yuanyuan{@Hanxian, this is the place holder for the over-an-over approach.} 
We first attempted a straightforward method of repeatedly forecasting the next $k$ queries using previously predicted $k$ queries as input for the next round of prediction. This process continues until there are enough predicted queries to cover the next $\Delta t$ interval. However, we empirically find that this approach results in poor accuracy, particularly as the number of prediction iterations increases. For instance, with $k=1000$, the 10 largest templates in the \wmc workload require 4 to 10 rounds of iterative predictions to cover a day's forecast, yielding an average accuracy of only $41.7\%$. As a result, we opted for a different approach.

\begin{figure}[t!]
% \vspace{-5pt}
  \centering
  \includegraphics[width=\linewidth]{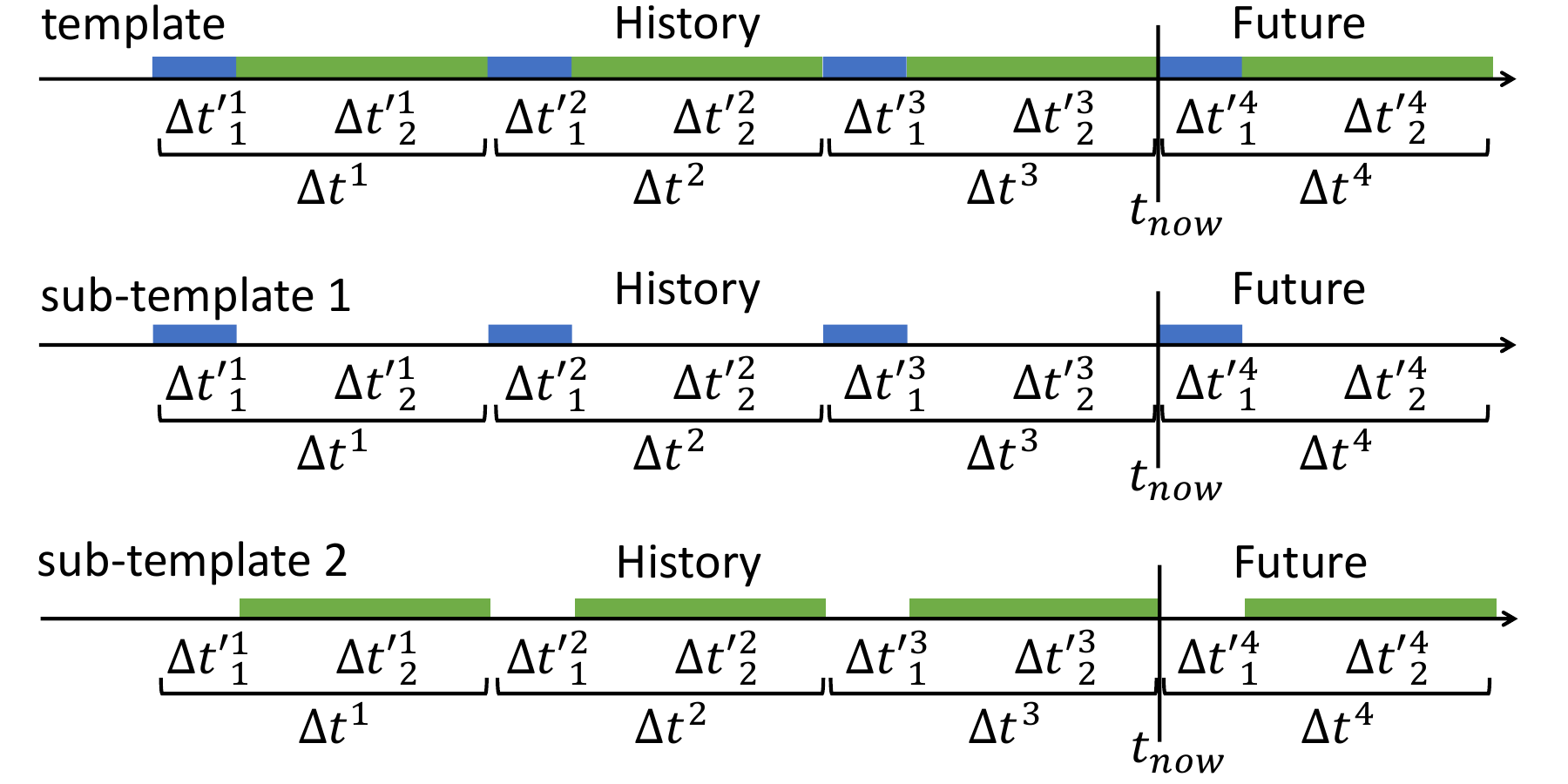}
%   \vspace{-4pt}
  \caption{Illustration of template cutting. %The blue bars correspond to queries that fall within the $\Delta t'_1$ timeframe of each $\Delta t$ interval, whereas the green bars correspond to queries that fall within the $\Delta t'_2$ timeframe of each $\Delta t$ interval. 
}
%\vspace{-15pt}
%\vspace{-3mm}
  \label{fig:cut}
\end{figure}

\revision{A continuous sub-sampling of a regular pattern in time series data still follows a regular pattern~\cite{oppenheim1999discrete}. This is because the sub-sampling process is essentially selecting a subset of the original pattern at regular intervals. As long as these intervals are consistent, the resulting pattern will also be regular.} \revision{Based on the above observation, we} propose template cutting to split the group of queries for a large template into sub-groups, \aka \textit{sub-templates}, so that each is no larger than $k$. We cut templates based on the arrival time, as illustrated in Figure~\ref{fig:cut}. Specifically, we divide the $\Delta t$ interval into \revision{consistent} sub-intervals, $\Delta t'_1,\Delta t'_2$  $, ..., \Delta t'_s$, and use them to divide the group of queries. Note that the sub-intervals do not have to be equal length in time. Figure~\ref{fig:cut} shows two sub-intervals $\Delta t'_1$ and $\Delta t'_2$ of $\Delta t$. After splitting, we now have two sub-templates. The first only contains queries that fall into the $\Delta t'_1$ time-frame of every $\Delta t$ interval (represented as blue bars in the figure), whereas the second contains queries in the $\Delta t'_2$ time-frame of every $\Delta t$ interval (represented as green bars). Assuming an accurate model for each sub-template, it will learn the new patterns (only having queries in its corresponding sub-interval) in the sub-template, and predict future queries based on the new patterns (only predicting queries in its corresponding sub-interval). 
%For instance, considering $\Delta t =$ 1 day, if $k<\sigma_i(1 \ day)$ but $k \geq \sigma_i(half \ day)$, we can split the template into two sub-templates: one with queries arriving between [0:00, 12:00) of a day and the other with queries arriving between [12:00, 24:00) of a day. %The first sub-template captures all morning queries of a day, while the second captures all afternoon queries of a day. 
%Intuitively, we can then treat each sub-template as a new template and train two separate per-template models for these two sub-templates, allowing each to learn the evolving patterns for its specific time interval. 
Combining the forecasted queries from both models will result in the forecasted queries for the entire $\Delta t$ interval.

For template cutting, we need to estimate the number of queries (see \S\ref{sec:tempalteSize}) that will arrive for a given template in next $\Delta t$, denoted as $\widetilde{\sigma_i}(\Delta t)$. %We discuss in %the next subsection 
%\S\ref{sec:tempalteSize} how we perform the estimation. 
Since we need to cut $\Delta t$ into smaller sub-intervals, we first choose a sufficiently fine-grained time window $\Delta t''$ to evenly divide $\Delta t$ into finer intervals $\Delta t''_1, \Delta t''_2, ..., \Delta t''_m$. For example, if $\Delta t$ = 1 day, we may pick $\Delta t''$ = 1 hour and have 24 small time intervals. We first use $\widetilde{\sigma_i}(\Delta t)$ to identify templates larger than $k$. For each such template, we examine the finer time granularity $\Delta t''$ and find the cutting points of sub-templates in the boundary of finer time intervals.
%We also choose a sufficiently fine-grained time window $\Delta t''$ of $\Delta t$ since we cut the interval into smaller sub-intervals. %that can be set based on the tuning interval at which performance is optimized. 
%For example, if $\Delta t$ = 1 day, we may pick $\Delta t'$ = 1 hour.
%We find that $\Delta t'$ = 1 hour is reasonably small for most tuning applications. \hanxian{this is not true. eg for auto tuning we have minute granualarity}
%We first use $\widetilde{\sigma_i}(\Delta t)$ to identify templates with sizes greater than $k$. For each such template, we examine the finer time granularity $\Delta t'$ and find the cutting points of sub-templates in the boundary of finer time intervals.
%The boundary of each finer time interval can be used as a cutting point of sub-templates. 
As depicted in Algorithm~\ref{algo:cutting}, it begins with $\Delta t''_1$ and greedily searches for a cutting point that results in the largest sub-template with a size no greater than $k$ by utilizing the estimated $\widetilde{\sigma_i}(\Delta t''_j)$. After the cut is made, it moves on iteratively to search for the cutting point for the next sub-template.

%\vspace{-2mm}
{
%\small
\begin{algorithm}
%\vspace{-1mm}
\caption{Template Cutting Algorithm}
\label{algo:cutting}
\begin{algorithmic}[1]
\State Input: $k$; $\widetilde{\sigma}_i(\Delta t''_j)$ and $\widetilde{\sigma}_i(\Delta t)$ for template $temp_i$. 
\State Initialize the final sub-template set $S=\emptyset$. 
\For {$temp_i$ in all templates}
    \State Initialize the sub-template set for $temp_i: s_i=\emptyset$
    \If {$\widetilde{\sigma}_i(\Delta t) > k$}  
        \State Current\_sub\_template  $C=\emptyset$
        \State Current\_remaining\_size $R=k$
        \For{$\Delta t''_j$ in $\Delta t$}
            \If  {$\widetilde{\sigma}_i(\Delta t''_j) \leq R$} 
                \State \textcolor{gray}{\# fit into the current sub-template}
                \State $C = C \cup \{temp_i(\Delta t''_j)\}, \ R=R-\widetilde{\sigma}_i(\Delta t'')$
            \Else
                \State \textcolor{gray}{\# initialize a new sub-template}
                \State $s_i = s_i \cup \{C\}$
                \State $C=\{temp_i(\Delta t''_j)\},R=k-\widetilde{\sigma}_i(\Delta t''_j)$                      
            \EndIf
        \EndFor 
    \Else
        \State $s_i=\{temp_i\}$
    \EndIf
    \State $S = S\cup s_i$
\EndFor
\State Output: the final sub-template set $S$ %, $s.t. k \geq \sum_{i \in s_j} \hat{\sigma}_i(\Delta t'), \forall s_j \in S$
\end{algorithmic}
%\vspace{-20pt}
\end{algorithm}
}

%\vspace{-4mm}
\subsection{Template Packing} \label{sec:pack}
%After cutting, the size of each template is no greater than $k$ (we treat each sub-template as a new template itself).
As mentioned in \textit{Challenge 2}, there are also many templates in the workload with very small sizes. We propose to `pack' multiple such smaller templates into bins such that the total number of expected queries per bin is no greater than $k$. We formulate template packing as an integer linear programming problem:

%\vspace{-10pt}

{%\small
%\footnotesize
\begin{equation*}
\begin{aligned}
& \text{minimize} \ \ &&\#bins = \sum_j y_j \\
& \text{subject to} 
    && \#bins \geq 1\\
&    &&  \sum_{i \in bin_j}(\widetilde{\sigma}_i(\Delta t)) \leq k, \ \forall temp_i \in bin_j, \ \forall bin_j  \\ %not exceeding k
&   && \#templates\_in\_bin_j \leq d,  \ \forall bin_j  \\
& && \sum_j x_{ij} =1, \forall temp_i \\ %assigning every template to exactly one bin
& && y_j, x_{ij} \in \{0,1\}, \forall temp_i, bin_j
\end{aligned}
\end{equation*}
}

\begin{comment} %this is a simpler version - but not super formal
{\small
\begin{equation*}
\begin{aligned}
& \text{minimize} \ \ &&\#bins \\
& \text{subject to} 
&&  \sum_i(\widetilde{\sigma}_i(\Delta t)) \leq k, \ \forall temp_i \in bin_j, \ \forall bin_j  \\
& && \#templates\_in\_bin_j \leq d,  \ \forall bin_j  
\end{aligned}
\end{equation*}
\end{comment}

Where $y_j=1$ if bin j is used and $x_{ij}=1$ if template $i$ is put into bin $j$. We modified a classical first-fit bin-packing algorithm~\cite{martello1990knapsack} to solve this problem by preferring a bin with fewer templates when multiple bins can fit a template. \revision{This is necessary because an excess of templates in a bin can affect the accuracy of the per-bin model due to the complex mixture of various template patterns. To mitigate this, we quantitatively set the bin size, or the maximum number of templates per bin, to a constant $d$.}

%When packing the small templates into bins, we also consider the number of templates per bin, \ie we avoid aggressively packing too many small templates into a bin.
%This is necessary because if there are too many templates to predict in a bin, the per-bin model's accuracy suffers due to the complex mixture of different template patterns. Quantitatively, we set the bin size, \ie the maximum number of templates per bin, to a constant $d$ to alleviate this problem. 
%The value of $d$ controls the degree of `packing' and brings a trade-off: a smaller $d$ may lead to a higher model accuracy but can result in more bins and thus more per-bin models to train and a higher overhead.  
%Empirically, we set $d=50$ in \sys, as it strikes a good balance between model accuracy and model management overhead for our workloads.

After packing, we can now create a model for each bin, which learns the distinct patterns for the templates present in the bin and predicts future queries for these templates. It is worth mentioning that, after packing, the bins are usually not fully occupied (\ie with sizes smaller than k), an effective model for the bin will accurately predict queries for these templates in the next $\Delta t$ interval, even with normal minor variations of template size in real workloads.

%Given that we ensure the total size of the templates is limited to $k$, an effective model for the bin will encompass queries for these templates in the next $\Delta t$ timeframe.

%It is worth mentioning that, after packing, the bins are usually not fully occupied (\ie with sizes smaller than k), which allows accurately forecasting even with normal minor variations of template size in real workloads. %\tar{do we evaluate the right size of d?}\hanxian{we did not evaluate d in S9. We just mentioned an empirical number here.}

%\vspace{-2.5mm}
\subsection{Per-Bin Models}\label{sec:perbinmodel}

We now adapt the per-template models into per-bin models to solve the next-$\Delta t$ forecasting problem. We still use the \sysModel as the forecasting models, but have to make the following adaptations:
\noindent
\textbf{The feature map} is more complex for per-bin model, since it includes queries from all templates within a bin. To handle multiple templates within each bin during featurization, it is necessary to include the template id as a feature and the feature map must also encompass the parameter values from all templates within the bin. \revision{We concatenate all the parameters from various templates in the feature map %(\eg a bin with two templates with 3 and 2 parameters respectively, will have a feature map with 5 parameters).
%(\eg a bin containing two templates, one with 3 parameters and the other with 2 parameters, will result in a feature map consisting of a total of 5 parameters).
(\eg two templates in a bin, with 3 and 2 parameters each, make a feature map of 5 parameters in the query feature vector).}

\noindent
\textbf{The forecasting task} becomes more challenging when modeling a mixed patterns from %multiple 
various templates in a bin. The per-bin model must accurately forecast template id and all parameter values within a bin for correctly reconstructing with the correct template and its parameter values.

%\item \textbf{The online feedback loop} is also improved. It is crucial to not only monitor model accuracy, fine-tune models, and detect new and inactive templates, but also to keep track of the size of each template. If the total size of templates exceeds the bin capacity, it becomes necessary to re-adjust the assignment of templates to bins and re-train the models for changed bins. 
%\squishend

%With template \cutNpack, we adjust the per-template models to per-bin models by adapting the following aspects: (1) In preprocessing, we first \cutNpack\ the templates and build one feature map for each bin. (2) Instead of using one feature map to encode one template, we expand the feature map to encode all the attributes of all the templates in a bin. (3) We include the template ID so that the model can also forecast which template ID will come next in a bin. (4) When reconstructing the forecasted query, we refer to the template context by the template ID, and then fill the corresponding parameters into the template. %When constructing the feature map, if one attribute appears in the template, the featurization is the same as discussed in Section~\ref{sec:featurization}. Otherwise we set it as a special categorical value $\$$, as shown in Figure~\ref{fig:per_cluster_temp2}
%We deploy the per-bin models to solve the next time interval forecasting problem. We identify the forecasted queries for the next time interval by the forecasted arrival time.

%\vspace{-2.5mm}
\subsection{Estimating Template Size}\label{sec:tempalteSize}
We now discuss how to estimate the template size used in the template cutting and packing.  %processes. 
To do so, we train an additional one-layer LSTM. We use the finer $\Delta t''$ granularity (introduced earlier) to collect $\sigma_i(\Delta t''_1), \sigma_i(\Delta t''_2),...,\sigma_i(\Delta t''_m)$ for each $\Delta t$ interval in the historical workload, and train the LSTM model to forecast the future arrival rates $\hat{\sigma_i}(\Delta t''_1), \hat{\sigma_i}(\Delta t''_2),..., \hat{\sigma_i}(\Delta t''_m)$. Summing up these predictions, we can %then 
compute $\hat{\sigma_i}(\Delta t)$.
%We collect $\sigma_i(\Delta t''_j)$ in the historical workload as a time series to train the LSTM model to forecast the future arrival rates in the $\Delta t''$ granularity (introduced earlier), denoted as $\hat{\sigma_i}(\Delta t'')$.  Using $\hat{\sigma_i}(\Delta t'')$, we can then compute $\hat{\sigma_i}(\Delta t)$. 
To ensure our next-$\Delta t$ forecasting models can stably predict multiple successive $\Delta t$ windows without costly retraining, we also set a longer forecasting horizon $\Delta T$, e.g., $\Delta T$ = 1 week given $\Delta t$ = 1 day. This provides a longer preview of future arrival rates, $\hat{\sigma_i}(\Delta t^1), \hat{\sigma_i}(\Delta t^2),..., \hat{\sigma_i}(\Delta t^l)$. We then conservatively use the upper-bound of all forecasted $\hat{\sigma_i}(\Delta t^j)$ to approximate the template size in next $\Delta t$, i.e., $\widetilde{\sigma_i}(\Delta t)=\max(\hat{\sigma}_i(\Delta t^j)), \forall \Delta t^j \in \Delta T$. Similarly, we can approximate the template size for smaller interval $\Delta t''_x$ %of $\Delta t$ 
as $\widetilde{\sigma_i}(\Delta t''_x)=\max(\hat{\sigma}_i(\Delta t''^j_x))$, $\forall \Delta t''^j \in \Delta t$.

Note that this LSTM model is only used to estimate the size of a template for template cutting and packing. The forecasted arrival time from \sysModel still determines the actual number of queries in the next-$\Delta t$ prediction. Note that while one-layer LSTM model is sufficient for our intended purpose, we can also use QueryBot 5000~\cite{ma2018query} to estimate the template size. 
\section{Feedback Loop}\label{sec:feedbackloop}
%\tar{I think we are not highlighting/selling this subsection enough in earlier parts of the paper...} \rana{Yuanyuan mentioned she is already adding this in the introduction }
Real workloads can shift, and new evolving patterns can emerge
that the pre-trained models have never seen, which leads to an accuracy degradation.
%Since real workload may dynamically change or shift, during the online stage, 
\sys\ offers a feedback loop to adapt to workload changes. Firstly, it tracks forecasting accuracy, detects changes in workloads, and automatically refines the models to enhance their performance. Secondly, it monitors both new and existing templates and keeps track of their sizes.

As \sys\ receives new queries continuously, it also receives the ground truth queries for the previous forecasting. This allows \sys to monitor the forecasting accuracy and decide whether to fine-tune the models or not. %Ideally, if the workload follows the time-evolving trend observed in the historical workload used to train our \sysModel models, the models can reliably generate consistently accurate predictions. 
%Real workloads can shift, and new evolving patterns can emerge
%that the pre-trained models have never seen, which leads to an accuracy degradation. 
To identify the workload shifts that trigger the accuracy degradation, we set an accuracy threshold. The threshold can be decided by the lower bound of the forecasting accuracy expectation by applications or DBA. In our study, we fine-tune a model if the model accuracy is constantly lower than the threshold $\alpha=75\%$ in a few forecasting rounds. %\tar{individual template/bin? or something else} \tar{over what window?}\hanxian{change to `model accuracy'}. %\tar{need to better motivate why 75\%. I will keep it a parameter here, and mention $75\%$ in experiments. I'm not sure if we can do any empirical evaluation now to justify this?}
\revision{During fine-tuning, as \sys collects new training data, new categorical values might emerge. We extend the dictionary of the parameter values by assigning new categorical values for them, and then use feature hashing to encode them.}

In addition to detecting pattern changes in existing query templates, \sys also has the capability to continuously identify emerging templates (\ie unseen templates that are recurrent in the new observation)
%\tar{define emerging templates}\tar{if something is seen only once, does it become emerging?} 
as well as inactive templates (\ie templates that have no queries showing up for a prolonged period of time). 
%It collects training data for the new templates, while removing the inactive templates from the models. 
For a new template, we collect training data for it while continuously receiving new queries. Then we either fit the new template into an existing bin, if the bin capacity allows it, and fine-tune the existing per-bin model, or otherwise initialize a new bin for it and train a new model on the collected training data. For an inactive template, since the template has no queries showing up, during periodic fine-tuning on the new observed data, the model will automatically not forecast queries for the template anymore.
Simultaneously, \sys keeps track of the size of each template and maintains models for the template size prediction (\S\ref{sec:tempalteSize}). If the total size of templates for a bin steadily exceeds the bin capacity, %the model forecasts a smaller number of queries than ground truth and leads to accuracy degradation. In this case, 
we divide the bin and re-adjust the assignment of templates into sub-bins so that the total size of templates in each sub-bin is no greater than $k$, and train new models for sub-bins from scratch. 
\section{Effectiveness Measurement}
\label{sec:accuracy}

%Measuring the accuracy of the forecasting results highly depends on the specific use case or the application for which we want to forecast the workload. 
As discussed in \S\ref{sec:Problem}, the majority of workload optimization tools that \sys targets at assume normal workloads instead of timed-workloads as input, so we only consider the normal workloads from the forecast when measuring prediction accuracy. 
%Below we discuss a general metric that can be customized for specific scenarios. 
Given the predicted workload $\hat{W}$ and the ground-truth future workload $W$, we use recall, precision, and F1 score as our evaluation metric with a customizable function 
$match(W, \hat{W})$ that defines how the ground-truth queries in $W$ are matched with the queries in $\hat{W}$, and can be tailored to a specific application. More formally, 

%\vspace{-10pt}
%\small
%\footnotesize
\begin{equation}\label{eq:recall}
    Recall = \frac{|match(W, \hat{W})|}{|W|}, \ \ \ \ Precision = \frac{|match(W, \hat{W})|}{|\hat{W}|}%, \ \ \ \ F1 = \frac{2 \cdot precision \cdot recall}{precision+recall}
    \nonumber
\end{equation}
%\begin{equation}\label{eq:precision}
%    Precision = \frac{|match(W, \hat{W})|}{|\hat{W}|} \nonumber
%\end{equation}
\begin{equation}\label{eq:f1}
    F1 = \frac{2 \cdot precision \cdot recall}{precision+recall} \nonumber
\end{equation}

%In the above equations, $match(Q, \hat{Q})$ denotes a correct forecasting identifier function returns $1$ if the forecasted query contains the groundtruth otherwise returns $0$. 
%When checking the \textit{match} function (i.e. multiset, allowing duplicates) of ground-truth queries that are matched to the predicted queries, and can be tailored to a specific application (details later).

%\iffalse
%A derived metric $F1$ score is the harmonic mean of these two metrics:
%\begin{equation}\label{eq:f1}
%    F1 = \frac{2 \cdot precision \cdot recall}{precision+recall}
%\end{equation}

%A high recall represents a high coverage of the ground truth. A high precision shows that the model returns more relevant results than irrelevant ones. 
%A higher $F1$ shows a higher prediction accuracy considering both recall and precision. In this paper, we use all three metrics to measure the accuracy of \sysModel models, while using various definitions of the prediction match function $match(Q, \hat{Q})$.
%\fi

While one can use a strict matching where the predicated queries exactly match with the ground truth queries, such a matching is rarely needed. For a large number of applications, such as index tuning, view recommendation, partitioning of tables, and determination of the MDC of tables, a forecasted workload is required cover most of the ground-truth queries, and a \textit{containment} based metric can be used for matching a ground-truth query with a predicted query.
%both parameter values as well as arrival time of forecasted queries exactly matches with the ground truth queries, such a matching is rarely needed. For a large number of applications, such as caching, index tuning, view recommendation, partitioning of tables, and determination of the MDC of tables, a forecasted workload for a specific period of time should cover most of the ground-truth queries. The precise ordering of queries or actual arrival time is insignificant in a targeting time window, rather 
%we want to only determine which queries should be included in given time window for which we perform tuning. Thus, the forecasted workload, in the given time window, can be treated as a bag of queries. Similarly, a \textit{containment} metric as described below can be used for matching a ground-truth query with a predicted query.
%A query $q_1$ is \textit{contained} in another query $q_2$, i.e., $q_1\subseteq q_2$, if for every database instance $d$, $q_1(d)\subseteq q_2(d)$, where $q_i(d)$ is the bag of tuples resulted in evaluating $q_i$ on $d$. 
In \sys, we measure containment only using predicates in the queries. For equality predicate, the model gives a single value, hence we perform exact matching. For range predicates, a match is considered if the predicted range contains the ground-truth range. In the case of an IN clause, a match is considered if the predicted value is a superset of the ground-truth value. 
Given the forecasted workload $\hat{W}$ and the ground-truth workload $W$, we use the containment relationship to define a bipartite graph $G$, where each query $q$ in $W$ and each query $\hat{q}$ in $\hat{W}$ serve as the vertices, and an edge exists between $q$ and $\hat{q}$, if $q$ is contained by $\hat{q}$. Then we define the containment-based match $match_{\subseteq}(W, \hat{W})$ as the maximum bipartite matching of $G$. Practically, we use a popular textbook greedy bipartite matching algorithm to approximate the optimal result. 

%we define the containment-based match as $match_{\subseteq}(W, \hat{W})=\{q\in W \mathrel{|}\exists \hat{q}\in \hat{W}, \mathrel{} q\subseteq \hat{q}\}$. 

To better understand the effectiveness of forecasting in terms of containment, we also want to measure the degree of the containment relationship for each \textit{matched} ground-truth and predicted query pair. We develop a new metric called \textit{average containment-diff ratio} (\textit{cnt-diff}) computed as follows. For a predicted range $\hat{R}$ and a ground-truth range $R$, the \textit{cnt-diff ratio} is defined as $\frac{|\hat{R}-R|}{|R|}$, where $-$ is the range difference operation. In cases of a half-bounded range predicate, such as $col>a$, all the observed parameter values related to $col$ are used to obtain the upper bound $max_{col}$ and lower bound $min_{col}$, and the range $(a, \infty)$ is changed to $(a, max_{col})$ before computing the %containment-diff 
cnt-diff ratio. Similarly, for IN-clause predicates, given a predicted set $\hat{S}$ and ground-truth set $S$, the cnt-diff ratio is defined as $\frac{|\hat{S}-S|}{|S|}$, where $-$ is the set difference operation. A good containment-based match should have a cnt-diff ratio close to zero. Finally, the average cnt-diff ratio is computed across all range predicates and IN-clause predicates for all the \textit{matched queries} (we do not compute cnt-diff ratio for unmatched queries). 

Finally, as discussed in \S\ref{sec:motivation}, sometimes there exist a small percentage of unpredictable parameters in a workload. %Although 
While the predicted values for these parameters will unlikely match the ground truth, they will still reflect the randomness of these parameters in the predicted workload. The workload optimization tools may be able to handle them some time. For example, in some view selections, a recurrent predicate with random parameter values will be ignored or converted into a group-by-column. For partitioning recommendation, random parameter values indicate either a non-ideal column for partitioning or a hash partitioning scheme for the column. An index recommender can take the randomness as a hint for needing an index structure better for point queries (e.g. hash-based indexes). \revision{Random parameters cannot be predicted. We believe the right approach is to be able to identify them and not apply any constraints on them.} We do not want to unnecessarily penalize a forecasting method for these unpredictable parameters. As a result, we do not consider the unpredictable parameters when reporting the accuracy measurements in this paper.

%\textbf{Handling unpredictable parameters.} Since we give entire query statements as part of the workload to the tuning applications, we need to provide values for unpredictable parameters (even though the percentage of such parameters is very small as discussed in \S\ref{sec:motivation}). 

%While handling such parameters depends on the applications, we propose the following approach that can be used for many tuning applications. Specifically, we generalize the corresponding predicate into a catch-all predicate, \textit{TRUE}. As an example, for the template \texttt{"SELECT ... FROM ... WHERE id = \$1 AND age > \$2 ..."}, if parameter \$1 is unpredictable, we transform its predicate \texttt{"id = \$1"} to \texttt{"TRUE"}, with predicted query as \texttt{"SELECT ... FROM ... WHERE TRUE AND age > 18 ..."}. By this generalization, we ensure that a forecasted query with accurate values for all the predictable parameters is a containment-based match to the ground truth.

% \section{Applications}\label{sec:apps}
% \rana{TODO}
% \subsection{View Materialization}

% %\subsection{Indexing}

% \subsection{Auto-Scaling}
% %with the forecasted workload, we (1) know the arrival rate of future workload + (2) can estimate the cost for each query better with the forecasted query --> auto-scaling 
%\vspace{-4mm}
\section{Evaluation}\label{sec:eval}

%\yuanyuan{@Hanxian, can you change historical-based to history-based in all the figures in this section?}

%We first evaluate the individual accuracy for each element in the feature vector and demonstrate our classifying on the predictable and unpredictable parameters. We then evaluate the query forecasting accuracy for the next-$k$ and the next-$\Delta t$ queries forecasting problems. We show the time and storage efficiency improvement by the \cutNpack. Then we give an example of finetuning on the shift workload. Finally, we demonstrate the real benefits gain of \sys\ with two DBMS applications: materialized view selection and semantic caching.

In this section, we study the parameter predictability (\S\ref{sec:individual_acc}), evaluate the effectiveness and efficiency of \sys for both %the 
next-$k$ (\S\ref{sec:nextK}) and next-$\Delta t$ forecasting problems (\S\ref{sec:nextDelta}), discuss the effectiveness of fine-tunning (\S\ref{sec:eval_fineTunning}), and demonstrate the real benefit gained by \sys for the view recommendation and index selection applications (\S\ref{sec:eval_applications}).

\textbf{Model Alternatives.}
We compare with three alternative models: RF, vanilla LSTM, and a heuristic \textit{history-based model}, which assumes a static workload and uses the last $k$ queries or the queries in the last $\Delta t$ window as the next-$k$ or next-$\Delta t$ forecast. %\tar{justify why these three? Explain why not CMU one?}

%\noindent\textbf{Historical-based:}
%A common non-ML based forecaster is to suggest the high-frequent historical queries with the assumption that the workload is static. For the next-$k$ and the next $\Delta t$ forecasting problems, we assume it recommends the historical $k$ queries and the queries in the last $\Delta t$ as the future queries accordingly.

%\noindent\textbf{Random Forest:} Random Forest is a ensemble learning method to construct one decision tree to deal with each subsample, and averege all the devisions from decision trees to be the final results.
%firsts splits the features by bootstrap sampling and construct one decision tree to deal with each subsample. Every decision tree makes its individual decision based on the subsample features and finally all the results from the decision trees are aggregated and averaged out to be the final result. 
%In our evaluation, we set the number of decision trees the same as the size of output window. 

%\noindent\textbf{Vanilla LSTM:}
%We adapt a vanilla LSTM model for the query forecasting problem. For the LSTM layer, we set the number of cells the same as the size of output window. 

\textbf{Implementation.} Workload templatization was implemented in Java using Calcite parser (v1.32.0)~\cite{DBLP:conf/sigmod/BegoliCHML18}. We implemented the rest of \sys\ in Python (v3.10.8) and built the ML models with Scikit-learn (v1.1.3)~\cite{Scikit-learn} and TensorFlow/Keras framework (v2.11.0)~\cite{abadi2016tensorflow}. For all ML models, we split each dataset such that the first 75\% of the sequence is used for training and the last 25\% for testing.  For RF, we set the number of decision trees the same as the output window size $k$. For Vanilla LSTM and \sysModel, we set the number of cells in each LSTM layer the same as the output window size $k$ and train with batches of $512$ samples until convergence or reaching the maximum number of training epochs $20$. We set the input window size equal to the output window size $k$ during model training and testing. We initialize the learning rate as $1e^{-3}$ with decay rate $0.9$, and use Adam optimization~\cite{kingma2014adam} and Huber loss~\cite{huber1992robust} implemented by TensorFlow/Keras. In our experiments, we use the accuracy threshold $\alpha=75\%$ to identify workload shift, \revision{and set the bin size $d=50$ for the per-bin models, unless otherwise specified}. %Whenever the model accuracy falls below 75\%, \sys triggers model fine-tuning.

\vspace{1mm}
\textbf{Experiment Setup.}
\revision{We used 6-core 206-GHz Xeon E5-2690V4 machines with Ubuntu 20.04 OS and one NVIDIA V100 GPU (16GB) for all experiments. 
%For all our experiments, we used machines equipped with a 6-core 206-GHz Xeon E5-2690V4 processor and running on Ubuntu 20.04 OS. Each machine is also fitted with one NVIDIA V100 GPU with 16GB memory that was utilized for ML training and testing. In \sys, 
Models for different templates or bins can be parallelized across machines to reduce the elapsed time. For inference, the models are indexed by their template or bin id, which allows loading only the required models at forecasting time.}

\subsection{Parameter Predictability}\label{sec:individual_acc}
\begin{figure}[th!]
%\vspace{-1pt}
  \centering
  \includegraphics[width=\linewidth]{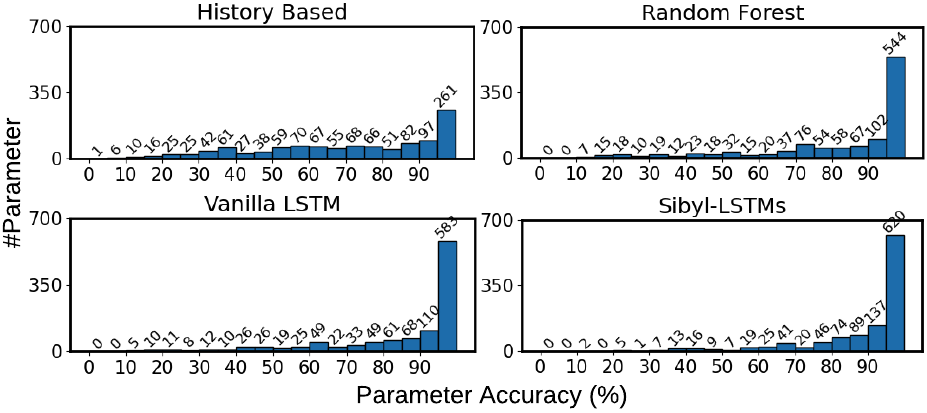}
  \caption{ Histogram of prediction accuracy for parameters in the \wmc workload, with forecasting window  $k=1000$. %\rana{increase the font size}
  }
  \label{fig:acc_hist}
%     \vspace{-11pt}
\end{figure}

\begin{figure}[th!]
%     \vspace{-10pt}
  \centering
  \includegraphics[width=\linewidth]{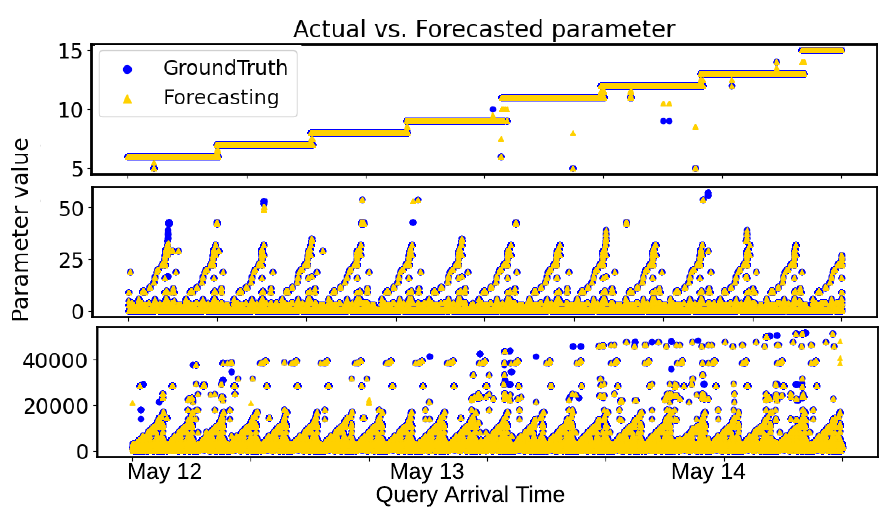}
  \caption{Forecasting visualization.%\rana{increase the size}
  }
  \label{fig:visualization}
%     \vspace{-13pt}
\end{figure}

%\tar{It's not totally clear what the set up here is. Why one template and not aggregate results over multiple templates? How is the histogram constructed from the predicted time series? Maybe worth clarifying these points.}
As discussed in \S\ref{sec:motivation}, although most parameters in time-evolving queries are predictable, a few exhibit random behavior. Figure~\ref{fig:acc_hist} details the %\hanxian{strict-match}\yuanyuan{I think we are fine here. Strict match is only for defining match of queries. Here, we are discussing the accuracy of parameters. So, we are just using the normal definition of accuracy here.} 
prediction accuracy histograms for all the parameters in the \wmc workload with the four different models in the next-1000 forecast. For each parameter, given the forecasted parameter sequence $\hat{p}_{n+1}, \hat{p}_{n+2}, ... , \hat{p}_{n+k}$ and the ground truth $p_{n+1}, p_{n+2}, ... , p_{n+k}$, the parameter accuracy is defined as $\frac{|\hat{p}_{n+i}=p_{n+i}|}{k}, \forall i \in [1,k]$. We can observe that more parameters can be accurately predicted using the ML-based models than the history-based model. Our \sysModel performs the best. 
%\yuanyuan{The rest of this paragraph and Figure~\ref{fig:visualization} could be commented out if we run out of space. } 
In Figure~\ref{fig:visualization}, \revision{we present a visualization of the actual versus forecasted values for the three predictable parameters (DATE, STRING, and ID-related fields, respectively) for one of the biggest templates in the \wmc workload, as an example.} We observed that the forecasted results closely aligned with the ground truth, particularly for periodic and trend patterns.

Figure~\ref{fig:acc_hist} %also 
shows that there are some unpredictable parameters for which even ML-based models cannot achieve over $75\%$ prediction accuracy. We visualized the time-series patterns for these parameters and observed the similar random patterns as shown in Figure~\ref{fig:CommonPatterns}(d). %To quantitatively identify unpredictable parameters, we introduce an accuracy threshold $\tau$, and treat parameters with accuracy lower than $\tau$ as unpredictable parameters. Users can customize $\tau$ according to the workloads and applications, as well as user expectation of accuracy. 
%In our evaluation, we empirically set $\tau=75\%$. 
For the remaining experiments, we ignore the unpredictable parameters as discussed in \S\ref{sec:accuracy}, and report accuracy using the containment-based match definition (also introduced in \S\ref{sec:accuracy}).%\hanxian{containment-based match for the next-k, containment-based+time-sensitive match for the next delta.}\yuanyuan{We are also fine here. We are actually only using the containment-based match. Time sensitive match does average of all accuracies in small time windows. But we are not. I am actually thinking about commenting out the time sensitive match altogether, since we never used it.}\hanxian{Agree}

%\vspace{-2mm}
\subsection{Next-$k$ Forecasting}
\label{sec:nextK}
We first show the forecasting accuracy of the per-template models for the next-k forecasting problem. We report the results on the time-evolving templates, ignoring the templates where the parameter values do not change over time.

%In our evaluation, we only report the results on the evolving parts of the workload , as shown in Table~\ref{tab:evolution}. The static parts of the workload are trivial to forecast using models.
\subsubsection{\textbf{Comparison of Models}}
\label{sec:kmodelcomparision}
Table~\ref{tab:next-k-acc} shows the average recall results for the per-template models in each workload with different window sizes $k$. We note that the \mssales workload has relatively a smaller number of queries, leading to smaller $k$ values%for this workload
. This is necessary to ensure sufficient training samples when sliding the input window over the query traces. %In the next-$k$ forecast problem, since $|W|=|\hat{W}|=k$, the recall, precision, and $F1$ score values are identical, resulting in a single reported accuracy value.
The results show that \sysModel clearly outperforms all the other models. The history-based approach has very low accuracy, especially for the \wmc, \scope and BusTracker workloads. Vanilla LSTM generally works better than the RF model. \revision{To ensure that the predicted queries do not overshoot for range and IN-clause predicates in the containment-based matches, we %also 
report the cnt-diff (see \S\ref{sec:accuracy}) for the forecast in Table~\ref{tab:proportional_diff}. Note that cnt-diff is calculated only on correct predictions. The results show that \sysModel often matches or surpasses cnt-diff ratios of other ML-based approaches, indicating that \sysModel does not attain the superior model accuracy by over-prediction.}

\subsubsection{\textbf{The Effect of $k$}}
The selection of $k$ depends on computational and memory resources available. We set the max forecasting window size as $1000$ to avoid the out-of-memory error given the machine memory constraint. As shown in Table~\ref{tab:next-k-acc}, the selection of $k$ also affects model accuracy. A smaller $k$ can result in a more accurate forecast but may also increase the risk of model instability or over-fitting. 
A larger $k$ predicts for a large forecasting window at once, but it poses challenges on accuracy and model scalability.
%A larger $k$ means more queries included in the training and prediction window, but this increases the model complexity and adds more strains on the computational resources.
%A larger $k$ brings a wider observation but also poses more challenges to the forecast-or models to digest more information in a longer-term memory. 
The experimental results show that \sysModel has better model scalability -- when scaling up the prediction window size, the variance of the accuracy is smaller compared to the other baselines. %\tar{we could add specific details on resource strains...} 

\Huge
\begin{table}[h!]
\caption{Accuracy results (\%) for next-$k$ forecasting problem. %on the four workloads. %\rana{increase the font size} %\yuanyuan{Please remove the last column}
}
\label{tab:next-k-acc}

\resizebox{\linewidth}{!}{
\begin{tabular}{|c|ccc|ccc|ccc|ccc|}
\hline
                 & \multicolumn{3}{c|}{\textbf{\wmc}}                                                                 & \multicolumn{3}{c|}{\textbf{\scope}}                                                              & \multicolumn{3}{c|}{\textbf{BusTracker}}                                                         & \multicolumn{3}{c|}{\textbf{\mssales}}                                                            \\ \hline
$k$                & \multicolumn{1}{c|}{100}           & \multicolumn{1}{c|}{500}           & 1000          & \multicolumn{1}{c|}{100}           & \multicolumn{1}{c|}{500}           & 1000          & \multicolumn{1}{c|}{100}           & \multicolumn{1}{c|}{500}           & 1000          & \multicolumn{1}{c|}{100}           & \multicolumn{1}{c|}{200}           & 500           \\ \hline
History-based & \multicolumn{1}{c|}{27.4}          & \multicolumn{1}{c|}{17.0}          & 31.8          & \multicolumn{1}{c|}{7.0}           & \multicolumn{1}{c|}{13.4}          & 32.8          & \multicolumn{1}{c|}{12.8}          & \multicolumn{1}{c|}{11.2}          & 7.9           & \multicolumn{1}{c|}{47.8}          & \multicolumn{1}{c|}{64.9}          & 72.8          \\
Random Forest    & \multicolumn{1}{c|}{85.4}          & \multicolumn{1}{c|}{82.3}          & 80.5          & \multicolumn{1}{c|}{83.7}          & \multicolumn{1}{c|}{83.2}          & 82.6          & \multicolumn{1}{c|}{91.4}          & \multicolumn{1}{c|}{90.2}          & 88.6          & \multicolumn{1}{c|}{79.6}          & \multicolumn{1}{c|}{75.3}          & 71.2          \\
Vanilla LSTM     & \multicolumn{1}{c|}{91.0}          & \multicolumn{1}{c|}{90.3}          & 90.1          & \multicolumn{1}{c|}{89.3}          & \multicolumn{1}{c|}{88.7}          & 88.2          & \multicolumn{1}{c|}{92.0}          & \multicolumn{1}{c|}{92.3}          & 91.8          & \multicolumn{1}{c|}{84.9}          & \multicolumn{1}{c|}{85.3}          & 80.7          \\
\sysModel\      & \multicolumn{1}{c|}{\textbf{95.8}} & \multicolumn{1}{c|}{\textbf{96.7}} & \textbf{95.4} & \multicolumn{1}{c|}{\textbf{94.6}} & \multicolumn{1}{c|}{\textbf{95.4}} & \textbf{94.7} & \multicolumn{1}{c|}{\textbf{96.0}} & \multicolumn{1}{c|}{\textbf{96.2}} & \textbf{95.8} & \multicolumn{1}{c|}{\textbf{92.4}} & \multicolumn{1}{c|}{\textbf{91.7}} & \textbf{88.2} \\ \hline
\end{tabular}}
%  \vspace{-5pt}
\end{table}

\begin{table}[h!]
  \caption{Cnt-diff ratio (\%) for the next-$k$ forecasting problem.% and the mean proportional difference across various $k$ settings and workloads for each model. %\rana{increase the font size}
  }
  \label{tab:proportional_diff}
    \resizebox{\linewidth}{!}{
\begin{tabular}{|c|ccc|ccc|ccc|ccc|}
\hline
 & \multicolumn{3}{c|}{\textbf{\wmc}}                                      & \multicolumn{3}{c|}{\textbf{\scope}}                                   & \multicolumn{3}{c|}{\textbf{BusTracker}}                              & \multicolumn{3}{c|}{\textbf{\mssales}}                                 \\ \hline
$k$ & \multicolumn{1}{c|}{100}  & \multicolumn{1}{c|}{500}  & 1000 & \multicolumn{1}{c|}{100}  & \multicolumn{1}{c|}{500}  & 1000 & \multicolumn{1}{c|}{100}  & \multicolumn{1}{c|}{500}  & 1000 & \multicolumn{1}{c|}{100}  & \multicolumn{1}{c|}{200}  & 500  \\ \hline
History-based                                                       & \multicolumn{1}{c|}{3.50} & \multicolumn{1}{c|}{4.24} & 1.01 & \multicolumn{1}{c|}{1.14} & \multicolumn{1}{c|}{1.19} & 0.73 & \multicolumn{1}{c|}{1.88} & \multicolumn{1}{c|}{2.86} & 3.04 & \multicolumn{1}{c|}{0.50} & \multicolumn{1}{c|}{0.49} & 0.55 \\
Random Forest                                                          & \multicolumn{1}{c|}{0.15} & \multicolumn{1}{c|}{0.04} & 0.16 & \multicolumn{1}{c|}{0.08} & \multicolumn{1}{c|}{0.08} & 0.12 & \multicolumn{1}{c|}{0.22} & \multicolumn{1}{c|}{0.11} & 0.22 & \multicolumn{1}{c|}{0.25} & \multicolumn{1}{c|}{0.09} & 0.28 \\
Vanilla LSTM                                                           & \multicolumn{1}{c|}{0.32} & \multicolumn{1}{c|}{0.11} & 0.21 & \multicolumn{1}{c|}{0.11} & \multicolumn{1}{c|}{0.18} & 0.07 & \multicolumn{1}{c|}{0.31} & \multicolumn{1}{c|}{0.31} & 0.35 & \multicolumn{1}{c|}{0.11} & \multicolumn{1}{c|}{0.19} & 0.04 \\
\sysModel\                                                           & \multicolumn{1}{c|}{0.19} & \multicolumn{1}{c|}{0.22} & 0.16 & \multicolumn{1}{c|}{0.16} & \multicolumn{1}{c|}{0.13} & 0.10 & \multicolumn{1}{c|}{0.25} & \multicolumn{1}{c|}{0.16} & 0.19 & \multicolumn{1}{c|}{0.36} & \multicolumn{1}{c|}{0.26} & 0.14 \\ \hline
\end{tabular}
}
%   \vspace{-3pt}
\end{table}

\normalsize

%\vspace{-2mm}
\subsection{Next-$\Delta t$ Forecasting}
\label{sec:nextDelta}
In Figure~\ref{fig:acc}, we present the evaluation results of per-bin models for the next-$\Delta t$ problem, varying $\Delta t$ to 1 hour, 6 hours, 12 hours, and 1 day. These time intervals are typical in our targeted application~\cite{jindal2018computation,jindal2021production}. %\tar{any references?}\hanxian{@Rana, could you help on this?}. 
We set $k$ (in template cutting and packing algorithms) to $1000$ for the \wmc, \scope, and BusTracker workloads, and $500$ for \mssales, as it has fewer queries.
%The template cutting and packing algorithms are used to organize templates into bins with a value of $k$ set to $1000$ for the \wmc, \scope, and BusTracker workloads, and $500$ for \mssales, as it has fewer queries. %We measure the recall, precision, and $F1$ score using the containment-based match definition.

%We set the input and output window sizes both $1000$ for \wmc, \scope, BusTracker and $500$ for \mssales since it has a relative smaller number of queries, so as to ensure sufficient training samples when input window sliding over the query traces. We then identify whether one query falls into the next $\Delta t$ according to the forecasted arrival time. In this case, usually $|Q| \neq |\hat{Q}|$. We report all three metrics recall, precision and $F1$ score, under the combined containment-based and time-sensitive accuracy definition, in Figure~\ref{fig:acc}.

\subsubsection{\textbf{Comparison of Models}}\label{subsec:compofmodels}
%Comparing various forecasting models, \sys significantly outperforms other methods and shows the most stable accuracy results with various $\Delta t$ settings. 
\sys %outperforms 
surpasses other forecasting models and maintains stable accuracy across various $\Delta t$ settings. Vanilla LSTM and Random Forecast perform poorly on the \mssales, which has more outliers and more unstable patterns. For \wmc, the history-based method performs well with the 12-hour interval due to the workload's recurrent queries that have the same parameter values within a day (between the past 12-hour window and the future 12-hour window). But this method is %not %as 
%effective 
ineffective with the one-day interval, as many query parameter values change when crossing the day boundary. The history-based method yields unsatisfactory results for the other three workloads that exhibit more rapid and intricate evolution and involve time-related parameters that operate on a finer time scale. Therefore, it is imperative to use an ML-based forecasting model to handle the evolving workload.

\begin{figure*}[t!]
%   \vspace{-10pt}
  \centering
  \includegraphics[width=\linewidth]{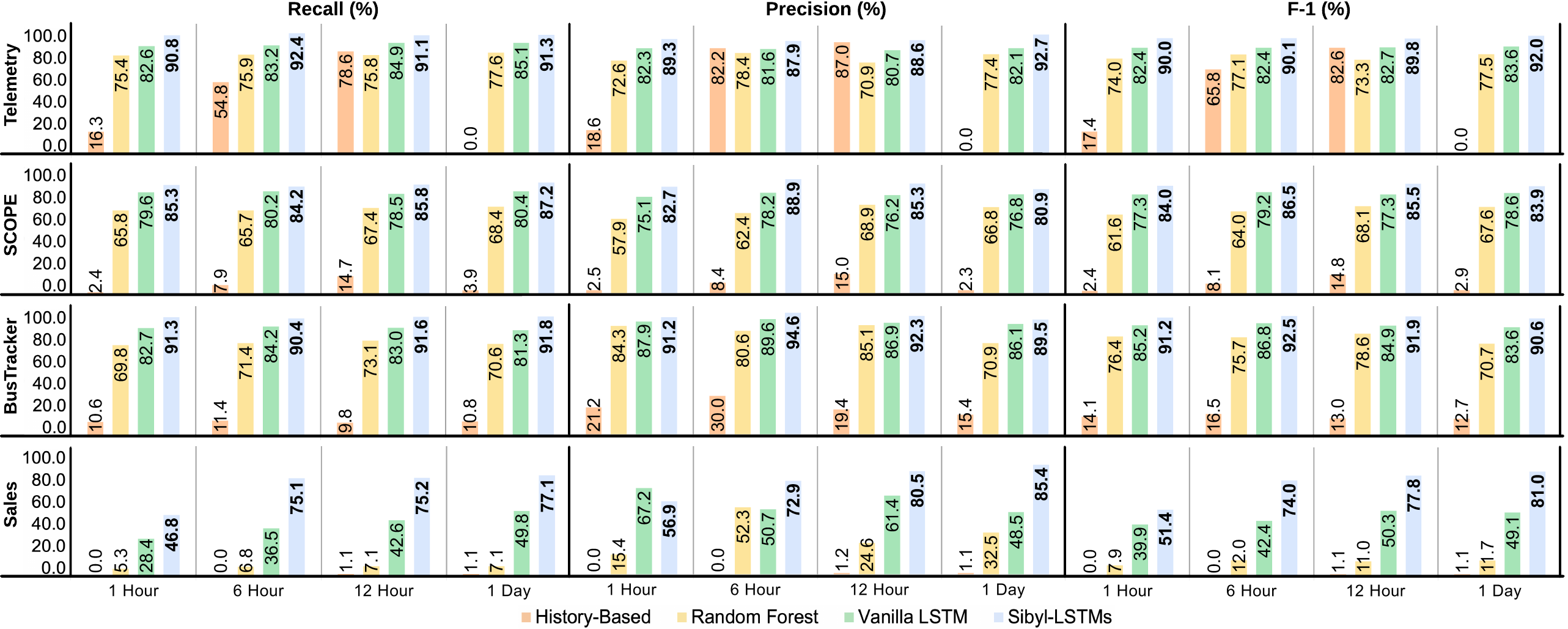}
  \caption{Accuracy results for the next-$\Delta t$ forecasting problem.}
  \label{fig:acc}
%     \vspace{-10pt}
\end{figure*}

\vspace{1mm}
\revision{\noindent\textbf{Other ML alternatives:} 
As noted earlier, our primary contribution lies in leveraging an effective ML model for our complex task, instead of developing new or exhaustively testing possible ML algorithms.
%We want to emphasize again that the paper's key contribution is applying ML to a complex database problem, not creating new ML algorithms. %While we've explored a number of ML models, others could also be applicable. 
%It’s not practical or even necessary to examine all possible models exhaustively; we only need one effective model. 
Here, we consider two more common alternatives although additional models are possible. (1) In 
another adaption of RF, referred to as RF+, we concatenate all values from $q_{n-k+1}$ to $q_{n}$ as a single input vector and output all parameters of $q_{n+1}$ to $q_{n+k}$ together. As Table~\ref{tab:CompMoreModels} shows, RF+ achieves similar low accuracy as the baseline RF in Figure~\ref{fig:acc}. (2) We also consider a Transformer based decoder-only model (TRF), with a 12-layer multi-head attention architecture. %, 440MB model size for each per-bin model %\yuanyuan{Do we want to mention the model size? Since 440MB is still much smaller than 54GB that we need for \sysModel.} 
%and \todox{} training time. 
Table~\ref{tab:CompMoreModels} shows TRF attains a slightly better accuracy than \sysModel, but at the cost of $18.2\times$ more training time and $4.4\times$ more model storage overhead%than \sysModel
. We chose \sysModel for \sys, because it offers a trade-off between accuracy and efficiency.
%a much higher price on the training overhead (97 minutes per model, $18.2\times$ than \sys) and model storage overhead (440MB per model, $4.4\times$ than \sys).
%We adjust Random Forest (RF) to use all the values from $q_{n-k+1}$ to $q_{n}$ as input by concatenating all the parameters in $q_{n-k+1}$ to $q_{n}$ into a single input vector, and get all parameters of $q_{n+1}$ to $q_{n+k}$ as output together. We refer such adaptation as RF+ and show the results in Table~\ref{tab:CompMoreModels}. RF+ shows lower forecasting accuracy than \sys by on average $15.2\%$ on various settings due to RF+'s limited ability to capture the sequential dependencies and handle the complex multi-variable multi-step forecasting.
%(2) We compare \sys with Transformer based decoder-only model, with a 12-layer multi-head attention architecture, with 440MB model size and ?? training time.
}
%\hanxian{I am thinking a SIBYL+, which is the cutting only next-k SIBYL (no packing), and it is more accurate and more efficient than TRF, if we have space for it.}

%\vspace{-2pt}
\begin{table}[h!]

\caption{\revision{The recall results (\%) comparison with other ML alternatives for the next-$\Delta t$ forecasting on the \wmc workload.}}
\label{tab:CompMoreModels}
%\vspace{-5pt}
\resizebox{0.7\linewidth}{!}{
\begin{threeparttable}
\begin{tabular}{|c|c|c|c|c|}
\hline
$\Delta t$ & 1 hour & 6 hour & 12 hour & 1 day \\ \hline
\revision{RF+} & \revision{75.5} & \revision{75.9} & \revision{75.8} & \revision{77.5} \\ \hline
\revision{TRF} & \revision{92.6 } & \revision{ 92.5} & \revision{ 93.0} & \revision{ 92.1} \\ \hline
\sys\        &    90.8    &    92.4    &     91.1    &    91.3   \\ \hline

\end{tabular}
\end{threeparttable}
 }
%   \vspace{-5pt}
\end{table}

%Because in \wmc, there are some query parameters related to the date of query arrival, which is changing in different days. Historical-based method gains poor results on the other three workloads that present more rapid and detailed evolution, and involve time-related parameters that operate on a finer time scale. Thus a ML-based forecasting model for the evolving workload is necessary.
\subsubsection{\textbf{The Effect of $\Delta t$}}
The accuracy of next-$\Delta t$ forecasting results is influenced not only by the model's ability to accurately forecast queries but also by its ability to accurately forecast arrival times. As the forecasting time window increases, the accuracy of the results changes. Typically, time-series forecasting with shorter horizons is easier and more accurate. However, predictions for smaller time granularity, such as 1 hour instead of 1 day, tend to be noisier and subject to greater fluctuations in query arrival rates per hour than per day, which makes forecasting arrival hours more challenging than forecasting arrival days. %\tar{it is not clear why noisier ...} 
As a result, the majority of forecasting accuracy results with a one-day time window are higher than other time-window settings. %\tar{the complexity and scalability challenges not coming out... how do we automatically decide the right granularity? The observation that 1 day granularity is mostly better goes against our earlier argument where we motivate a need for cutting, etc.}

\subsubsection{\textbf{Time and Storage Efficiency}}\label{sec:efficiency}
Table~\ref{tab:estimatedPerModel} demonstrates the time and storage savings achieved through the implementation of template cutting and packing as well as the per-bin models for next-day forecasting. \revision{Note that the training times reported in the table are aggregation across all templates/bins. By parallelizing on multiple machines, the elapsed training times can be significantly shortened.} %The prediction overhead was not included in our analysis here since it typically takes only a few milliseconds to seconds to perform a forecasting on a GPU machine, which is trivial compared to training overhead.
We note that the average time and storage overhead of a single per-bin model is higher than that of a per-template model due to larger model capacity and a higher average number of queries per bin than per template. However, template cutting and packing significantly reduce the number of %total 
models by up to $23\times$. Moreover, employing per-bin models results in a significant reduction in training time of up to $13.6\times$ and storage space by up to $6.0\times$ when compared to per-template models. 

We  note that there is a trade-off between efficiency and accuracy. Comparing the accuracy results in Table~\ref{tab:next-k-acc} and Figure~\ref{fig:acc}, the accuracy of per-bin models is slightly lower than per-template models. Because the next-$\Delta t$ forecasting is a harder problem to solve than the next-$k$ forecasting, as mentioned in \S\ref{sec:perbinmodel}. It has a higher requirement for the per-bin models to forecast the query arrival time precisely, depending on which we can identify the queries in the next time interval correctly. Different from the per-template models, the per-bin models are required to forecast template ids in a bin and more queries in per bin than per template.

\revision{Compared to the training overhead, the total per-bin model prediction times on GPU are negligible: 3.9s, 241s, 1.6s, and 0.031s for \wmc, \scope, Bustracker, and \mssales, respectively.}

%\hanxian{All the training and prediction are conducted offline to prepare inputs for optimization applications.}

\Huge
\begin{table}[tbh!]
\centering
\caption{\revision{The aggregate training time} and model storage overhead for per-template and per-bin models. $\downarrow$ means the reduction ratio.%\rana{increase the font size} %\yuanyuan{@Hanxian, can you fill in this table?}
}
\label{tab:estimatedPerModel}

\resizebox{\linewidth}{!}{
\begin{threeparttable}

\begin{tabular}{c|c|c|c|c}
\hline
             & \wmc      & \scope     & BusTracker & \mssales \\ \hline
\# per-template models          &2157 &168197 &258 &23            \\ %\hline
\revision{aggregate} training time\tnote{$\dag$} \ (h) &119.8  & 9344.3   & 14.3          &    1.3           \\ %\hline
total model storage\tnote{$\dag$} \ (GB) & 54.0   & 4205.0  & 6.5         &       0.6      \\ 
\hline 
\hline
\# per-bin models ($\downarrow$) & 124 ($17.4\times$) & 7716 ($21.8\times$) & 50 ($5.2\times$) &  1 ($23\times$)          \\ %\hline
\revision{aggregate} training time\tnote{$\dag$} \ (h) \ ($\downarrow$)& 11.0  ($10.9\times$) & 685.9 ($13.6\times$)&    4.4 ($3.3\times$)    & 0.1 ($13.0\times$)              \\ %\hline
total model storage\tnote{$\dag$} \ (GB) \ ($\downarrow$)&  12.4  ($4.4\times$) & 771.6  ($5.4\times$)  &    5.0  ($1.3\times$)     &       0.1 ($6.0\times$)       \\ \hline
\end{tabular}
 \begin{tablenotes}
    %\footnotesize
    \huge
    \item[$\dag$] The one-epoch average training time is $10s$ for a per-template model and $16s$ for a per-bin model. The average storage overhead is $25MB$ for a per-template model and $100MB$ for a per-bin model. %The total training time and model storage are estimated by linearly scaling up.
 
  \end{tablenotes}

\end{threeparttable}
 }
%   \vspace{-5pt}
\end{table}
\normalsize

\subsubsection{\textbf{Comparison with Previous Work}}\label{sec:baselines}

We now compare \sys\ with QueryBot5000~\cite{ma2018query}\revision{, TEALED~\cite{tealed} \footnote{\revision{It is our re-implementation because the code/executable of TEALED is not available.}}} and Q-Learning~\cite{meduri2021evaluation} on the common BusTracker workload. Although \revision{none of the three methods were originally designed} for future query forecasting, we adapted them to solve the next-$\Delta t$ forecasting problem. QueryBot5000 \revision{and TEALED} forecast only the arrival rates for templates in the next $\Delta t$. To generate queries, we take the most recent $n$ historical queries from each template, where $n$ is the predicted query rate by QueryBot5000 \revision{and TEALED}. Q-Learning only forecasts the next \textit{one} query. We adapt it by continuously forecasting using the last predicted query as input until collecting $m$ queries, where $m$ is the number of queries in the last $\Delta t$. We call the adapted methods as QueryBot5000+, \revision{TEALED+} and Q-Learning+, respectively.

%a bag of queries with the same query count as the last $\Delta t$.
%we modified it by incorporating the recent historical queries from the clusters with forecasted query volumes as the next-$\Delta t$ forecasting results.

%and we adapt it by taking the recent historical queries of the clusters with the forecasted query volumes as its next-$\Delta t$ forecasting results. Meduri et al. leverage Q-Learning to forecast the next query, and we adapt it by continuously forecasting using the last forecasted query as input until collecting a bag of queries with the same query count as the last $\Delta t$. 
We compare the recall results \revision{and prediction overhead for next-1 day forecasting} in Table~\ref{tab:CompRelated}. \sys significantly outperforms QueryBot5000+ as the latter fails to forecast the time-evolving parameter values. QueryBot5000+ \revision{and TEALED+} show slightly better accuracy than the history-based method shown in Figure~\ref{fig:acc}, as they accurately forecast the query arrival rate. Q-Learning+ only suggests a similar query to the ground truth and has very low recall results when applying the over-and-over prediction. \revision{Q-Learning+ also has the highest prediction overhead because it forecasts only one query at a time, while other methods have comparable low prediction overhead.}
%\yuanyuan{@Hanxian: Are the prediction times for 1-day?}

%\vspace{-2pt}
\begin{table}[h!]
\caption{The recall results (\%) \revision{and prediction overhead} comparison for the next-$\Delta t$ forecasting on %the 
BusTracker.} %workload.}
\label{tab:CompRelated}
\resizebox{\linewidth}{!}{
\begin{threeparttable}
\begin{tabular}{|c|c|c|c|c|c|}
\hline
$\Delta t$ & 1 hour & 6 hour & 12 hour & 1 day  & %prediction overhead (1 day)\\
\multicolumn{1}{c|}{\begin{tabular}[c]{@{}c@{}}\revision{prediction}\\ \revision{overhead (1 day)}\end{tabular}}\\ 
\hline
QueryBot5000+      &    12.4    &    11.7    &     12.0    &    13.9    &\revision{1.1s}\\ \hline
\revision{TEALED+}& \revision {11.0}&\revision {11.5} & \revision {10.6}& \revision {12.0}&\revision {5.2s}\\\hline
Q-Learning+ & 0.0 & 0.0 & 0.0 & 0.0  & \revision{6251.4s} \\ \hline 
 \hline
\sys\        &    91.3    &    90.4    &     91.6    &    91.8    &\revision{1.6s} \\ \hline

\end{tabular}

 %\begin{tablenotes}
    %\footnotesize
    %\item[$\dag$] \revision{It is our re-implementation because the original TEALED is not open-sourced.}
  %\end{tablenotes}

\end{threeparttable}
 }
%\vspace{-5pt}   
\end{table}

\subsubsection{\textbf{\revision{{The Effect of Bin Size}}}}
\label{sec:effect_bin_size}
%As mentioned in \S~\ref{sec:pack}, we set a threshold $d$ to limit the bin size, \ie the maximum number of templates per bin to avoid aggressively packing and accuracy degradation. 
\revision{We now assess the impact of bin size $d$ (the maximum number of templates per bin) on the recall results of next-1 day forecasting on the \wmc workload, in Table~\ref{tab:binsize}. While per-template models ($d$=1) provide high accuracy, they are not time and storage efficient as discussed in \S~\ref{sec:efficiency}. Larger $d$ values reduce the number of models needed but compromise accuracy due to the complex mixture of patterns. We empirically set $d=50$ in \sys %for a good trade-off between 
to balance accuracy and efficiency.}

%Here we evaluate the effect of the bin size $d$, \ie the maximum number of templates per bin, on the recall results of the next-1 day forecasting on the \wmc workload, in Table~\ref{tab:binsize}. The per-template models ($d=1$) gain highest accuracy but they are not time and storage efficient as discussed in \S~\ref{sec:efficiency}. When a larger $d$ is set, there are more templates packed in a bin and thus fewer models needed to train for a given workload, but it also trades model accuracy off due to the more complicated patterns mixed of more templates. Starting from $d\simeq 50$, as the number of bin increasing, the number of bins and the recall results do not change too much, since only those very small templates are re-arranged in different bins, which have little effect on final recall results. In \sys, we empirically set $d=50$ since it achieves a good balance between model accuracy and overhead (\S~\ref{sec:efficiency}) for our workloads.
%\vspace{-3pt}
\begin{table}[ht!]
\caption{ \revision{The recall results (\%) of different $d$ settings for the next-1 day forecasting on the \wmc workload.}}
\label{tab:binsize}
\resizebox{0.75\linewidth}{!}{
\begin{tabular}{|c|c|c|c|c|c|c|c|}
\hline
\revision{d}      & \revision{1}    & \revision{10}   & \revision{30}   & \revision{50}   & \revision{70}   & \revision{100}\\ \hline
\revision{\#bins} & \revision{2157} & \revision{442}  & \revision{217}  & \revision{124}  & \revision{120}  & \revision{119}\\ \hline
\revision{Recall} & \revision{93.2} & \revision{92.0} & \revision{91.6} & \revision{91.3} & \revision{91.1} & \revision{91.1}\\ \hline

\end{tabular}}
%\vspace{-5pt}
\end{table}

%\subsection{Scalability}
%We evaluate the scalability of model by the variance on $F1$ score when scaling-up prediction window size.
%\vspace{-2mm}
\subsection{Effectiveness of Fine-tuning Models}
\label{sec:eval_fineTunning}
%As mentioned in \S\ref{sec:feedbackloop}, \sys\ monitors the forecasting accuracy, detects workload shift, and automatically finetunes the model to improve model accuracy. Here we show the detection of workload shift and finetuning, with a parameter in \wmc for the next-$1000$ forecasting as an example, in Figure~\ref{fig:finetune}. In Figure~\ref{fig:finetune}(a), the period of the a parameter pattern shortens starting from May 13 (marked in light blue) and \sys\ detects it by observing the accuracy degradation. The parameter accuracy is $51.3\%$ on the shifted pattern and lower than the threshold $\tau$, which triggers model finetuning. As shown in Figure~\ref{fig:finetune}(b), \sys\ fine-tunes the \sysModel\ by incrementally training on the new observed data, instead of training from scratch. The fine-tuning achieves convergent in only two epochs with only a few seconds fine-tuning overhead, and gains accuracy to $92.7\%$ which is close to the pre-train accuracy $93.0\%$.

\begin{figure}[h!]
%\vspace{-6pt}
  \centering
  \includegraphics[width=\linewidth]{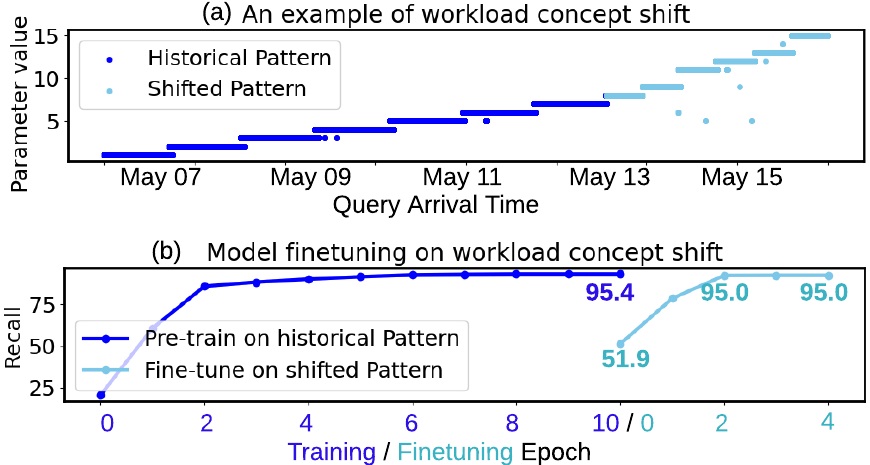}
  \caption{%\vspace{-2mm}
  Fine-tuning on \wmc workload shift. %\vspace{-6pt} %\rana{increase the font size}
  }
  \label{fig:finetune}
%  \vspace{-10pt}
\end{figure}

%In \S\ref{sec:feedbackloop}, we discussed \sys's feedback loop %how \sys\ continuously monitors forecasting accuracy, detects workload shifts, and automatically performs model finetuning 
In this experiment, we demonstrate the effectiveness of  \sys's feedback loop (see \S\ref{sec:feedbackloop}).
%to enhance accuracy.
Figure~\ref{fig:finetune} displays the detection of workload shift in the \wmc workload and the execution of a per-bin model fine-tuning for the next-day forecasting. Figure~\ref{fig:finetune}(a) depicts a pattern change of a parameter in the \wmc workload starting from May 13 (highlighted in light blue), which \sys\ detects by observing the decline in accuracy. The model accuracy on the shifted pattern is $51.9\%$, which falls below the threshold $\alpha=75\%$, triggering model fine-tuning. In Figure~\ref{fig:finetune}(b), we observe that \sys fine-tunes the \sysModel by incrementally training on newly observed data, rather than training from scratch. 
\revision{
The average fine-tuning time per epoch is under 10s for a per-template model and 16s for a per-bin model, due to the smaller size of new observed data. In \sys, we limit the maximum number of fine-tuning epochs to 20. As shown in Figure~\ref{fig:finetune}, the model converges in just two epochs with 6.4 seconds of overhead, improving accuracy to $95.0\%$, which is close to the pre-trained accuracy of $95.4\%$.}

%The one-epoch average fine-tuning time is less than 10s for a per-template model, and 16s for a per-bin model, since fine-tuning uses smaller-size new observed data than training. We set the max number of fine-tuning epochs as 20. In the example shown in Figure~\ref{fig:finetune}, the model converges on the shifted data in only two epochs with 6.4 seconds of overhead, and the accuracy improves to $95.0\%$, which is close to the pre-trained accuracy of $95.4\%$.
%Fine-tuning converges in only two epochs with a few seconds of overhead
%\yuanyuan{We mentioned a few minutes overhead in the author feedback. Is that the same thing? If so, why is here a few seconds?}
%, and the accuracy improves to $95.0\%$, which is close to the pre-trained accuracy of $95.4\%$.

\subsection{\revision{Applications of Workload Forecasting}}

\revision{We now show how workload forecast can be applied to two classical workload-based optimization applications: view selection and index selection. 
The purpose of our experiments is to show how an existing view/index selection algorithm can directly use the forecasted workload without modification to its algorithm to produce better views/indexes, rather than introducing a new view/index selection algorithm. Thus, we employ the well-known view selection algorithm~\cite{DBLP:conf/vldb/AgrawalCN00} and the PostgreSQL index recommender tool~\cite{pgindex}.} % implemented using Calcite~\cite{DBLP:conf/sigmod/BegoliCHML18}. 

\subsubsection{\textbf{Application to View Selection}}
\label{sec:eval_applications}
%To evaluate the effectiveness of our forecasting techniques in real-world scenarios, we apply them in two practical applications: \emph{materialized view selection} and \emph{semantic caching}. 

%\subsubsection{\textbf{Materialized View Selection}}
%\rana{Add more info about DW pool and size of base tables}\yuanyuan{@Rana, please fill in the numbers in the place holders.} \rana{Done!}

%Our evaluation uses a traditional materialized view algorithm as described in~\cite{DBLP:conf/vldb/AgrawalCN00}, implemented relying on Apache Calcite~\cite{DBLP:conf/sigmod/BegoliCHML18}. In a nutshell, this algorithm ($i$)~employs heuristics to prune a set of SPJA materialized views that are syntactically relevant to the given workload, and ($ii$)~efficiently searches through the remaining candidates to produce a final recommendation. 

\begin{figure}[th!]
%\vspace{-2mm}
 \centering
 \includegraphics[width=\columnwidth]{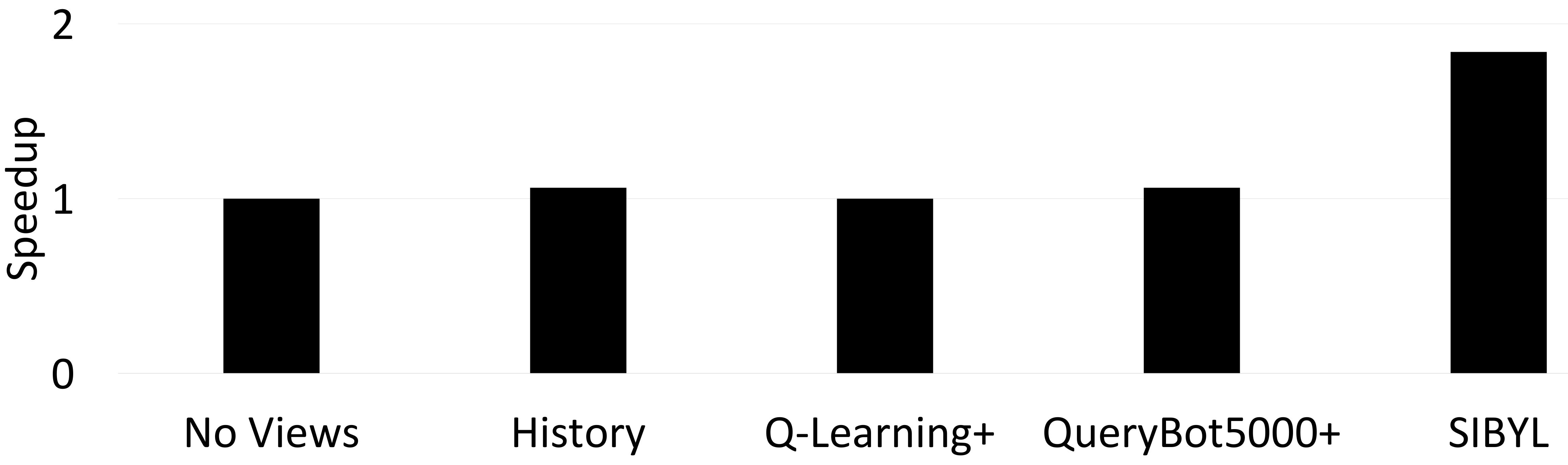}
 \caption{\revision{%\vspace{-2mm}
 Speedup via views on \mssales workload.}   %\rana{increase the font size}
 }
 \label{fig:views}
\end{figure}

\revision{We train \sysModel, QueryBot5000+, and Q-Learning+ using 2237 \mssales queries over 20 consecutive days. Then, we employ the view selection algorithm to create materialized views for the subsequent day. As the baseline, we conduct view recommendation on the preceding 7 days of \textit{history} queries. For \sysModel, QueryBot5000+, and Q-Learning+, we use the predicted queries to recommend views. Then we run the ground-truth queries using the recommended views on a cloud-based data warehouse (2-compute nodes and 385GB data). Figure~\ref{fig:views} shows the query execution speedup achieved by using the views compared to without the views. The queries forecasted by Q-Learning+ do not lead to any useful views at all, so there is no speedup. Views based on both History and QueryBot5000+ result in merely $1.06\times$ improvement, whereas views recommended based on \sys leads to $1.83\times$ speedup, roughly a $1.7\times$ difference.}

\begin{comment}
\begin{figure}[h]
\centering
\begin{minipage}{.5\linewidth}
  \centering
  \includegraphics[width=0.99\linewidth]{figure/views_v2.pdf}
  \caption{\revision{%\vspace{-2mm}
 Speedup via views on \mssales workload.}  }
  \label{fig:views}
\end{minipage}%
\hfill
\begin{minipage}{.5\linewidth}
\centering
  \includegraphics[width=0.99\linewidth]{figure/IndexResults.pdf}
  \caption{\revision{%\vspace{-2mm}
 Speedup via indexes on \wmc workload.}}
  \label{fig:indices}
\end{minipage}
\end{figure}
\end{comment}

%\vspace{-5pt}

\subsubsection{\revision{\textbf{Application to Index Selection}}}
\label{sec:eval_index_selection}
\begin{figure}[th!]
%\vspace{-2mm}
 \centering
 \includegraphics[width=\columnwidth]{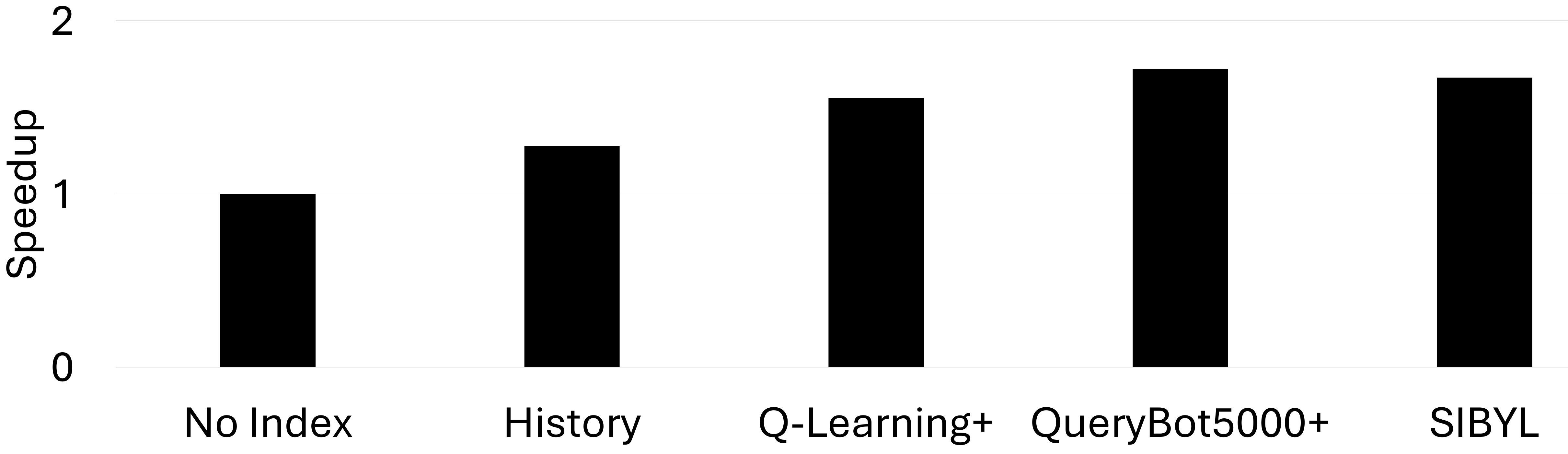}
 \caption{\revision{%\vspace{-2mm}
 Speedup via indexes on \wmc workload.}   %\rana{increase the font size}
 }
 \label{fig:indices}
\end{figure}

\revision{In this experiment, we train \sysModel, QueryBot5000+, and Q-Learning+ using 741K queries from 11 days of \wmc workload. We then run the index recommender for Day 12. Due to the large volumes of queries ($\approx$ 47K) on the 12th day, we only focus on the 151 queries that fall in the 3AM - 4AM window. For the baseline, we run the index recommender on a random sample of 1K queries from the historical workload, following the same approach in~\cite{ma2018query}. For \sysModel, QueryBot5000+, and Q-Learning+, we recommend indexes on the forecasted queries. Then, we execute the ground-truth queries on the recommended indexes using a single-node PostgreSQL server on 24GB data. Figure~\ref{fig:indices} shows the speedup achieved by various methods. \textit{History} exhibits a modest $1.2\times$ speedup when compared to \textit{No Index}. All three ML-based methods outperform \textit{History}. Not surprisingly, QueryBot5000+ achieves a very good $1.72\times$ speedup, since index recommendation is one of the target applications it is designed for~\cite{ma2018query}. \sys achieves a comparable $1.67\times$ speedup. It is important to note that the accuracy requirements for query prediction are less stringent for index recommendation. As long as the predicted queries encompass the main tables and columns, the recommended index will be beneficial for future workloads. In contrast, view recommendation necessitates precise prediction of query templates and parameter values to generate useful views.}

\section{Related Work}\label{sec:related}
% Workload modeling and forecasting have been extensively studied in the context of relational databases. The related work can be classified into three %broad 
% categories: ($i$) partial query forecasting, ($ii$) forecasting workload features, and ($iii$) suggesting queries from history. 

There has been extensive research on workload modeling and forecasting for relational databases, which can be classified into three main categories: ($i$) partial query forecasting, ($ii$) workload feature forecasting, and ($iii$) suggesting queries from the past.

\vspace{1mm}
\noindent\textbf{Partial Query Forecasting}. %The most related two works are~\cite{ma2018query} and~\cite{meduri2021evaluation} and the comparison is shown in Table~\ref{tab:comparison}. 
%\cite{meduri2021evaluation,jain2018query2vec} also predict future query workloads by encoding the workload and learning it by ML models. 
%Query2Vec~
\cite{jain2018query2vec} %proposes learning 
learns vector representations for %raw 
SQL statements and query plans, which captures the syntax similarity among query statements but fails to predict the literals in queries. %Meduri et al.~
\cite{meduri2021evaluation} leverage RNNs~\cite{medsker2001recurrent} and Q-Learning~
\cite{watkins1992q} to predict the next query based on the current query, with the forecasted literals as a bin of values rather than accurate values. While these works partially forecast query statements, \sys\ accurately predicts the entire future statements. 

\vspace{1mm}
\noindent\textbf{Forecasting Workload Features.} %Holze et al.~
\cite{holze2008autonomic} utilizes Markov models to predict the shifts in the workload over time, %while Holze et al.~\cite{holze2010towards}  
and~\cite{holze2010towards} models periodic %and recurring 
patterns of a workload by classification. %Prophet~
\cite{taylor2018forecasting} forecasts accessing frequency by ensemble learning and %QueryBot 5000~
\cite{ma2018query} predicts the arrival rate of future queries by hybrid-ensemble learning to suggest indexing. %Abebe et al.~
\cite{abebe2022tiresias} predicts data access characteristics such as latency, when data will be accessed, and volume of data accessed. \revision{The paper \cite{tealed} utilizes a combination of time-sensitive empirical mode decomposition (EMD) and
auto LSTM encoder-decoder to forecast resource utilization and query arrival rates for DBMSs.} These efforts focus on specific aspects of workload forecasting. In contrast, \sys\ %is a workload forecasting framework that not only forecasts future query statements and their arrival time but can also infer the query arrival rate and pattern shifts simultaneously. 
is a comprehensive workload forecasting framework that predicts query statements, arrival time, arrival rate, and pattern shifts simultaneously.

\vspace{1mm}
\noindent\textbf{Suggesting Queries From the Past}. Query recommendation~\cite{eirinaki2013querie,eirinaki2015querie} selects queries from the historical logs that overlap with ongoing interaction sessions using collaborative filtering. Query auto-completion~\cite{khoussainova2010snipsuggest} helps users complete the missing parts of a query by choosing transitions based on heuristics, such as the popularity of query fragment co-occurrence in prior logs. % (such as popularity of query fragment co-occurrence in prior logs and foreign key dependencies). The highest ranked transition (DAG edge) is then chosen to identify the complete query (child DAG node) from a given partial query (current DAG node). 
% These works rely on heuristics and historical statistics, and fail to suggest new queries that do not already exist in the history.
\revision{The paper \cite{8781017} posits that database workloads are influenced by real-world events. It forecasts future workloads by identifying upcoming events and matching them with similar past events. 
These approaches are not ML-based, but only rely on heuristics and historical workloads. They also do not capture query evolution, thus fail to suggest new queries that are not already in the history.}

%\noindent\textbf{Encoder-decoder architecture} has been widely adopted and shown success in sequence-to-sequence learning problems in many fields. The encoder-decoder architecture is originally brought up to solve the machine translation problem~\cite{cho2014learning}, then is adapted in many other applications, e.g., speech recognition~\cite{lu2015study,lu2016training}, vehicle trajectory prediction~\cite{park2018sequence,deo2018convolutional}, object detection~\cite{ji2021cnn}. In \sys, we formulate the workload forecasting problem as a sequence-to-sequence learning problem and adapt the encoder-decoder architecture with advanced stacked-LSTM layers to perform precise forecasting. %In recent years, there are increasingly fast-emerging large language models~\cite{vaswani2017attention,devlin-etal-2019-bert,radford2019language}. These models usually have more model parameters and take longer training time. They are less efficient than LSTM-based encoder-decoder architectures but usually have a good performance on complex sequence-to-sequence learning problem. Evaluating the advanced NLP solutions for our query forecasting problem on more complicated workloads is left for our future work.

%\tar{compare with Khuzaima's paper and briefly. Also add the citation in intro where you mention forecasting techniques. in intro we do not need to compare maybe.}\hanxian{todo: add "Prophet" ~\cite{taylor2018forecasting}}
\vspace{10mm}
\section{Conclusion}\label{sec:conclusion}

We introduced \sys, an ML-based workload forecasting framework that predicts future queries across various time intervals. Unlike the prior work,  \sys formulates the forecasting problem as a multi-variate, multi-step, sequence-to-sequence prediction problem. We addressed several challenges to efficiently and accurately predict future queries, which lead to performance improvement in applications such as views and indexes selection, emphasizing \sys's potential in database optimization. As future work, we plan to explore other ML techniques to further reduce the training overhead and improve efficiency.

\bibliographystyle{ACM-Reference-Format}
\bibliography{sample}


\begin{thebibliography}{56}

%%% the Web address.

\ifx \showCODEN    \undefined \def \showCODEN     #1{\unskip}     \fi
\ifx \showDOI      \undefined \def \showDOI       #1{#1}\fi
\ifx \showISBNx    \undefined \def \showISBNx     #1{\unskip}     \fi
\ifx \showISBNxiii \undefined \def \showISBNxiii  #1{\unskip}     \fi
\ifx \showISSN     \undefined \def \showISSN      #1{\unskip}     \fi
\ifx \showLCCN     \undefined \def \showLCCN      #1{\unskip}     \fi
\ifx \shownote     \undefined \def \shownote      #1{#1}          \fi
\ifx \showarticletitle \undefined \def \showarticletitle #1{#1}   \fi
\ifx \showURL      \undefined \def \showURL       {\relax}        \fi
% invisible to TeX
\providecommand\bibfield[2]{#2}
\providecommand\bibinfo[2]{#2}
\providecommand\natexlab[1]{#1}
\providecommand\showeprint[2][]{arXiv:#2}

\bibitem[pgi({[n.\,d.]})]%
        {pgindex}
 \bibinfo{year}{[n.\,d.]}\natexlab{}.
\newblock \bibinfo{title}{{Dexter}}.
\newblock \bibinfo{howpublished}{\url{https://github.com/ankane/dexter}}.
\newblock


\bibitem[db2({[n.\,d.]})]%
        {db2}
 \bibinfo{year}{[n.\,d.]}\natexlab{}.
\newblock \bibinfo{title}{{IBM Db2}}.
\newblock \bibinfo{howpublished}{\url{https://www.ibm.com/analytics/us/en/db2}}.
\newblock


\bibitem[inf({[n.\,d.]})]%
        {informix}
 \bibinfo{year}{[n.\,d.]}\natexlab{}.
\newblock \bibinfo{title}{{IBM Informix}}.
\newblock \bibinfo{howpublished}{\url{https://www.ibm.com/products/informix}}.
\newblock


\bibitem[sql({[n.\,d.]})]%
        {sqlserver}
 \bibinfo{year}{[n.\,d.]}\natexlab{}.
\newblock \bibinfo{title}{{Microsoft SQL Server}}.
\newblock \bibinfo{howpublished}{\url{https://www.microsoft.com/en-us/sql-server/sql-server-2022}}.
\newblock


\bibitem[ora({[n.\,d.]})]%
        {oracle}
 \bibinfo{year}{[n.\,d.]}\natexlab{}.
\newblock \bibinfo{title}{{Oracle}}.
\newblock \bibinfo{howpublished}{\url{https://www.oracle.com/database}}.
\newblock


\bibitem[par({[n.\,d.]})]%
        {parametermarkers}
 \bibinfo{year}{[n.\,d.]}\natexlab{}.
\newblock \bibinfo{title}{{SQL Server - Parameter Markers}}.
\newblock \bibinfo{howpublished}{\url{https://learn.microsoft.com/sql/odbc/reference/appendixes/parameter-markers}}.
\newblock


\bibitem[Abadi et~al\mbox{.}(2016)]%
        {abadi2016tensorflow}
\bibfield{author}{\bibinfo{person}{Mart{\'\i}n Abadi}, \bibinfo{person}{Paul Barham}, \bibinfo{person}{Jianmin Chen}, \bibinfo{person}{Zhifeng Chen}, \bibinfo{person}{Andy Davis}, \bibinfo{person}{Jeffrey Dean}, \bibinfo{person}{Matthieu Devin}, \bibinfo{person}{Sanjay Ghemawat}, \bibinfo{person}{Geoffrey Irving}, \bibinfo{person}{Michael Isard}, {et~al\mbox{.}}} \bibinfo{year}{2016}\natexlab{}.
\newblock \showarticletitle{Tensorflow: a system for large-scale machine learning.}. In \bibinfo{booktitle}{\emph{Osdi}}, Vol.~\bibinfo{volume}{16}. Savannah, GA, USA, \bibinfo{pages}{265--283}.
\newblock


\bibitem[Abebe et~al\mbox{.}(2022)]%
        {abebe2022tiresias}
\bibfield{author}{\bibinfo{person}{Michael Abebe}, \bibinfo{person}{Horatiu Lazu}, {and} \bibinfo{person}{Khuzaima Daudjee}.} \bibinfo{year}{2022}\natexlab{}.
\newblock \showarticletitle{Tiresias: enabling predictive autonomous storage and indexing}.
\newblock \bibinfo{journal}{\emph{Proceedings of the VLDB Endowment}} \bibinfo{volume}{15}, \bibinfo{number}{11} (\bibinfo{year}{2022}), \bibinfo{pages}{3126--3136}.
\newblock


\bibitem[Agrawal et~al\mbox{.}(2000)]%
        {DBLP:conf/vldb/AgrawalCN00}
\bibfield{author}{\bibinfo{person}{Sanjay Agrawal}, \bibinfo{person}{Surajit Chaudhuri}, {and} \bibinfo{person}{Vivek~R. Narasayya}.} \bibinfo{year}{2000}\natexlab{}.
\newblock \showarticletitle{Automated Selection of Materialized Views and Indexes in {SQL} Databases}. In \bibinfo{booktitle}{\emph{{VLDB} 2000, Proceedings of 26th International Conference on Very Large Data Bases, September 10-14, 2000, Cairo, Egypt}}. \bibinfo{publisher}{Morgan Kaufmann}, \bibinfo{pages}{496--505}.
\newblock


\bibitem[Begoli et~al\mbox{.}(2018)]%
        {DBLP:conf/sigmod/BegoliCHML18}
\bibfield{author}{\bibinfo{person}{Edmon Begoli}, \bibinfo{person}{Jes{\'{u}}s Camacho{-}Rodr{\'{\i}}guez}, \bibinfo{person}{Julian Hyde}, \bibinfo{person}{Michael~J. Mior}, {and} \bibinfo{person}{Daniel Lemire}.} \bibinfo{year}{2018}\natexlab{}.
\newblock \showarticletitle{Apache Calcite: {A} Foundational Framework for Optimized Query Processing Over Heterogeneous Data Sources}. In \bibinfo{booktitle}{\emph{Proceedings of the 2018 International Conference on Management of Data, {SIGMOD} Conference 2018, Houston, TX, USA, June 10-15, 2018}}. \bibinfo{publisher}{{ACM}}, \bibinfo{pages}{221--230}.
\newblock


\bibitem[Bruno et~al\mbox{.}(2011)]%
        {autoadmin}
\bibfield{author}{\bibinfo{person}{Nicolas Bruno}, \bibinfo{person}{Surajit Chaudhuri}, \bibinfo{person}{Arnd~Christian K{\"{o}}nig}, \bibinfo{person}{Vivek~R. Narasayya}, \bibinfo{person}{Ravishankar Ramamurthy}, {and} \bibinfo{person}{Manoj Syamala}.} \bibinfo{year}{2011}\natexlab{}.
\newblock \showarticletitle{AutoAdmin Project at Microsoft Research: Lessons Learned}.
\newblock \bibinfo{journal}{\emph{{IEEE} Data Eng. Bull.}} \bibinfo{volume}{34}, \bibinfo{number}{4} (\bibinfo{year}{2011}), \bibinfo{pages}{12--19}.
\newblock


\bibitem[Chaudhuri and Narasayya(2007)]%
        {selftune}
\bibfield{author}{\bibinfo{person}{Surajit Chaudhuri} {and} \bibinfo{person}{Vivek Narasayya}.} \bibinfo{year}{2007}\natexlab{}.
\newblock \showarticletitle{Self-Tuning Database Systems: A Decade of Progress}. In \bibinfo{booktitle}{\emph{VLDB '07}}. \bibinfo{pages}{3–14}.
\newblock


\bibitem[Cho et~al\mbox{.}(2014)]%
        {cho2014learning}
\bibfield{author}{\bibinfo{person}{Kyunghyun Cho}, \bibinfo{person}{Bart Van~Merri{\"e}nboer}, \bibinfo{person}{Caglar Gulcehre}, \bibinfo{person}{Dzmitry Bahdanau}, \bibinfo{person}{Fethi Bougares}, \bibinfo{person}{Holger Schwenk}, {and} \bibinfo{person}{Yoshua Bengio}.} \bibinfo{year}{2014}\natexlab{}.
\newblock \showarticletitle{Learning phrase representations using RNN encoder-decoder for statistical machine translation}.
\newblock \bibinfo{journal}{\emph{arXiv preprint arXiv:1406.1078}} (\bibinfo{year}{2014}).
\newblock


\bibitem[Chuckravanen({[n.\,d.]})]%
        {chuckravanen2014approximate}
\bibfield{author}{\bibinfo{person}{Dineshen Chuckravanen}.} \bibinfo{year}{[n.\,d.]}\natexlab{}.
\newblock \showarticletitle{Approximate entropy as a measure of cognitive fatigue: an eeg pilot study}.
\newblock  (\bibinfo{year}{[n.\,d.]}).
\newblock


\bibitem[Devlin et~al\mbox{.}(2019)]%
        {devlin-etal-2019-bert}
\bibfield{author}{\bibinfo{person}{Jacob Devlin}, \bibinfo{person}{Ming-Wei Chang}, \bibinfo{person}{Kenton Lee}, {and} \bibinfo{person}{Kristina Toutanova}.} \bibinfo{year}{2019}\natexlab{}.
\newblock \showarticletitle{{BERT}: Pre-training of Deep Bidirectional Transformers for Language Understanding}. In \bibinfo{booktitle}{\emph{Proceedings of the 2019 Conference of the North {A}merican Chapter of the Association for Computational Linguistics: Human Language Technologies, Volume 1 (Long and Short Papers)}}. \bibinfo{publisher}{Association for Computational Linguistics}, \bibinfo{address}{Minneapolis, Minnesota}, \bibinfo{pages}{4171--4186}.
\newblock
\urldef\tempurl%
\url{https://doi.org/10.18653/v1/N19-1423}
\showDOI{\tempurl}


\bibitem[{Edgar Haren}(2017)]%
        {oracleselfdrive}
\bibfield{author}{\bibinfo{person}{{Edgar Haren}}.} \bibinfo{year}{2017}\natexlab{}.
\newblock \bibinfo{title}{{Oracle Revolutionizes Cloud with the World’s First Self-Driving Database}}.
\newblock \bibinfo{howpublished}{\url{https://blogs.oracle.com/database/post/oracle-revolutionizes-cloud-with-the-worlds-first-self-driving-database}}.
\newblock


\bibitem[Eirinaki et~al\mbox{.}(2013)]%
        {eirinaki2013querie}
\bibfield{author}{\bibinfo{person}{Magdalini Eirinaki}, \bibinfo{person}{Suju Abraham}, \bibinfo{person}{Neoklis Polyzotis}, {and} \bibinfo{person}{Naushin Shaikh}.} \bibinfo{year}{2013}\natexlab{}.
\newblock \showarticletitle{Querie: Collaborative database exploration}.
\newblock \bibinfo{journal}{\emph{IEEE Transactions on knowledge and data engineering}} \bibinfo{volume}{26}, \bibinfo{number}{7} (\bibinfo{year}{2013}), \bibinfo{pages}{1778--1790}.
\newblock


\bibitem[Eirinaki and Patel(2015)]%
        {eirinaki2015querie}
\bibfield{author}{\bibinfo{person}{Magdalini Eirinaki} {and} \bibinfo{person}{Sweta Patel}.} \bibinfo{year}{2015}\natexlab{}.
\newblock \showarticletitle{QueRIE reloaded: Using matrix factorization to improve database query recommendations}. In \bibinfo{booktitle}{\emph{2015 IEEE International Conference on Big Data (Big Data)}}. IEEE, \bibinfo{pages}{1500--1508}.
\newblock


\bibitem[Getta(2018)]%
        {8781017}
\bibfield{author}{\bibinfo{person}{Janusz~R. Getta}.} \bibinfo{year}{2018}\natexlab{}.
\newblock \showarticletitle{Event Based Forecasting of Database Workloads}. In \bibinfo{booktitle}{\emph{2018 IEEE 4th International Conference on Computer and Communications (ICCC)}}. \bibinfo{pages}{1767--1773}.
\newblock


\bibitem[Hochreiter and Schmidhuber(1997)]%
        {hochreiter1997long}
\bibfield{author}{\bibinfo{person}{Sepp Hochreiter} {and} \bibinfo{person}{J{\"u}rgen Schmidhuber}.} \bibinfo{year}{1997}\natexlab{}.
\newblock \showarticletitle{Long short-term memory}.
\newblock \bibinfo{journal}{\emph{Neural computation}} \bibinfo{volume}{9}, \bibinfo{number}{8} (\bibinfo{year}{1997}), \bibinfo{pages}{1735--1780}.
\newblock


\bibitem[Holze et~al\mbox{.}(2010)]%
        {holze2010towards}
\bibfield{author}{\bibinfo{person}{Marc Holze}, \bibinfo{person}{Ali Haschimi}, {and} \bibinfo{person}{Norbert Ritter}.} \bibinfo{year}{2010}\natexlab{}.
\newblock \showarticletitle{Towards workload-aware self-management: Predicting significant workload shifts}. In \bibinfo{booktitle}{\emph{2010 IEEE 26th International Conference on Data Engineering Workshops (ICDEW 2010)}}. IEEE, \bibinfo{pages}{111--116}.
\newblock


\bibitem[Holze and Ritter(2008)]%
        {holze2008autonomic}
\bibfield{author}{\bibinfo{person}{Marc Holze} {and} \bibinfo{person}{Norbert Ritter}.} \bibinfo{year}{2008}\natexlab{}.
\newblock \showarticletitle{Autonomic databases: Detection of workload shifts with n-gram-models}. In \bibinfo{booktitle}{\emph{Advances in Databases and Information Systems: 12th East European Conference, ADBIS 2008, Pori, Finland, September 5-9, 2008. Proceedings 12}}. Springer, \bibinfo{pages}{127--142}.
\newblock


\bibitem[Huang et~al\mbox{.}(2022)]%
        {tealed}
\bibfield{author}{\bibinfo{person}{Xiuqi Huang}, \bibinfo{person}{Yunlong Cheng}, \bibinfo{person}{Xiaofeng Gao}, {and} \bibinfo{person}{Guihai Chen}.} \bibinfo{year}{2022}\natexlab{}.
\newblock \showarticletitle{TEALED: A Multi-Step Workload Forecasting Approach Using Time-Sensitive EMD and Auto LSTM Encoder-Decoder}. In \bibinfo{booktitle}{\emph{Database Systems for Advanced Applications}}. \bibinfo{pages}{706--713}.
\newblock


\bibitem[Huber(1992)]%
        {huber1992robust}
\bibfield{author}{\bibinfo{person}{Peter~J Huber}.} \bibinfo{year}{1992}\natexlab{}.
\newblock \showarticletitle{Robust estimation of a location parameter}.
\newblock \bibinfo{journal}{\emph{Breakthroughs in statistics: Methodology and distribution}} (\bibinfo{year}{1992}), \bibinfo{pages}{492--518}.
\newblock


\bibitem[Jain et~al\mbox{.}(2018)]%
        {jain2018query2vec}
\bibfield{author}{\bibinfo{person}{Shrainik Jain}, \bibinfo{person}{Bill Howe}, \bibinfo{person}{Jiaqi Yan}, {and} \bibinfo{person}{Thierry Cruanes}.} \bibinfo{year}{2018}\natexlab{}.
\newblock \showarticletitle{Query2vec: An evaluation of NLP techniques for generalized workload analytics}.
\newblock \bibinfo{journal}{\emph{arXiv preprint arXiv:1801.05613}} (\bibinfo{year}{2018}).
\newblock


\bibitem[Jindal et~al\mbox{.}(2018a)]%
        {jindal2018selecting}
\bibfield{author}{\bibinfo{person}{Alekh Jindal}, \bibinfo{person}{Konstantinos Karanasos}, \bibinfo{person}{Sriram Rao}, {and} \bibinfo{person}{Hiren Patel}.} \bibinfo{year}{2018}\natexlab{a}.
\newblock \showarticletitle{Selecting subexpressions to materialize at datacenter scale}.
\newblock \bibinfo{journal}{\emph{{VLDB}}} \bibinfo{volume}{11}, \bibinfo{number}{7} (\bibinfo{year}{2018}), \bibinfo{pages}{800--812}.
\newblock


\bibitem[Jindal et~al\mbox{.}(2021)]%
        {jindal2021production}
\bibfield{author}{\bibinfo{person}{Alekh Jindal}, \bibinfo{person}{Shi Qiao}, \bibinfo{person}{Hiren Patel}, \bibinfo{person}{Abhishek Roy}, \bibinfo{person}{Jyoti Leeka}, {and} \bibinfo{person}{Brandon Haynes}.} \bibinfo{year}{2021}\natexlab{}.
\newblock \showarticletitle{Production Experiences from Computation Reuse at Microsoft.}. In \bibinfo{booktitle}{\emph{EDBT}}. \bibinfo{pages}{623--634}.
\newblock


\bibitem[Jindal et~al\mbox{.}(2018b)]%
        {jindal2018computation}
\bibfield{author}{\bibinfo{person}{Alekh Jindal}, \bibinfo{person}{Shi Qiao}, \bibinfo{person}{Hiren Patel}, \bibinfo{person}{Zhicheng Yin}, \bibinfo{person}{Jieming Di}, \bibinfo{person}{Malay Bag}, \bibinfo{person}{Marc Friedman}, \bibinfo{person}{Yifung Lin}, \bibinfo{person}{Konstantinos Karanasos}, {and} \bibinfo{person}{Sriram Rao}.} \bibinfo{year}{2018}\natexlab{b}.
\newblock \showarticletitle{Computation reuse in analytics job service at microsoft}. In \bibinfo{booktitle}{\emph{Proceedings of the 2018 International Conference on Management of Data}}. \bibinfo{pages}{191--203}.
\newblock


\bibitem[Khoussainova et~al\mbox{.}(2010)]%
        {khoussainova2010snipsuggest}
\bibfield{author}{\bibinfo{person}{Nodira Khoussainova}, \bibinfo{person}{YongChul Kwon}, \bibinfo{person}{Magdalena Balazinska}, {and} \bibinfo{person}{Dan Suciu}.} \bibinfo{year}{2010}\natexlab{}.
\newblock \showarticletitle{SnipSuggest: Context-aware autocompletion for SQL}.
\newblock \bibinfo{journal}{\emph{Proceedings of the VLDB Endowment}} \bibinfo{volume}{4}, \bibinfo{number}{1} (\bibinfo{year}{2010}), \bibinfo{pages}{22--33}.
\newblock


\bibitem[Kingma and Ba(2014)]%
        {kingma2014adam}
\bibfield{author}{\bibinfo{person}{Diederik~P Kingma} {and} \bibinfo{person}{Jimmy Ba}.} \bibinfo{year}{2014}\natexlab{}.
\newblock \showarticletitle{Adam: A method for stochastic optimization}.
\newblock \bibinfo{journal}{\emph{arXiv preprint arXiv:1412.6980}} (\bibinfo{year}{2014}).
\newblock


\bibitem[Li et~al\mbox{.}(2016)]%
        {li2016estimation}
\bibfield{author}{\bibinfo{person}{Xiaoling Li}, \bibinfo{person}{Ying Jiang}, \bibinfo{person}{Jun Hong}, \bibinfo{person}{Yuanzhe Dong}, {and} \bibinfo{person}{Lei Yao}.} \bibinfo{year}{2016}\natexlab{}.
\newblock \showarticletitle{Estimation of cognitive workload by approximate entropy of EEG}.
\newblock \bibinfo{journal}{\emph{Journal of Mechanics in Medicine and Biology}} \bibinfo{volume}{16}, \bibinfo{number}{06} (\bibinfo{year}{2016}), \bibinfo{pages}{1650077}.
\newblock


\bibitem[Lu et~al\mbox{.}(2016)]%
        {lu2016training}
\bibfield{author}{\bibinfo{person}{Liang Lu}, \bibinfo{person}{Xingxing Zhang}, {and} \bibinfo{person}{Steve Renais}.} \bibinfo{year}{2016}\natexlab{}.
\newblock \showarticletitle{On training the recurrent neural network encoder-decoder for large vocabulary end-to-end speech recognition}. In \bibinfo{booktitle}{\emph{2016 IEEE International Conference on Acoustics, Speech and Signal Processing (ICASSP)}}. IEEE, \bibinfo{pages}{5060--5064}.
\newblock


\bibitem[Ma et~al\mbox{.}(2018)]%
        {ma2018query}
\bibfield{author}{\bibinfo{person}{Lin Ma}, \bibinfo{person}{Dana Van~Aken}, \bibinfo{person}{Ahmed Hefny}, \bibinfo{person}{Gustavo Mezerhane}, \bibinfo{person}{Andrew Pavlo}, {and} \bibinfo{person}{Geoffrey~J Gordon}.} \bibinfo{year}{2018}\natexlab{}.
\newblock \showarticletitle{Query-based workload forecasting for self-driving database management systems}. In \bibinfo{booktitle}{\emph{Proceedings of the 2018 International Conference on Management of Data}}. \bibinfo{pages}{631--645}.
\newblock


\bibitem[Marcus et~al\mbox{.}(2022)]%
        {marcus2022bao}
\bibfield{author}{\bibinfo{person}{Ryan Marcus}, \bibinfo{person}{Parimarjan Negi}, \bibinfo{person}{Hongzi Mao}, \bibinfo{person}{Nesime Tatbul}, \bibinfo{person}{Mohammad Alizadeh}, {and} \bibinfo{person}{Tim Kraska}.} \bibinfo{year}{2022}\natexlab{}.
\newblock \showarticletitle{Bao: Making learned query optimization practical}.
\newblock \bibinfo{journal}{\emph{ACM SIGMOD Record}} \bibinfo{volume}{51}, \bibinfo{number}{1} (\bibinfo{year}{2022}), \bibinfo{pages}{6--13}.
\newblock


\bibitem[Marcus et~al\mbox{.}(2019)]%
        {marcus2019neo}
\bibfield{author}{\bibinfo{person}{Ryan Marcus}, \bibinfo{person}{Parimarjan Negi}, \bibinfo{person}{Hongzi Mao}, \bibinfo{person}{Chi Zhang}, \bibinfo{person}{Mohammad Alizadeh}, \bibinfo{person}{Tim Kraska}, \bibinfo{person}{Olga Papaemmanouil}, {and} \bibinfo{person}{Nesime Tatbul}.} \bibinfo{year}{2019}\natexlab{}.
\newblock \showarticletitle{Neo: A learned query optimizer}.
\newblock \bibinfo{journal}{\emph{arXiv preprint arXiv:1904.03711}} (\bibinfo{year}{2019}).
\newblock


\bibitem[Martello and Toth(1990)]%
        {martello1990knapsack}
\bibfield{author}{\bibinfo{person}{Silvano Martello} {and} \bibinfo{person}{Paolo Toth}.} \bibinfo{year}{1990}\natexlab{}.
\newblock \bibinfo{booktitle}{\emph{Knapsack problems: algorithms and computer implementations}}.
\newblock \bibinfo{publisher}{John Wiley \& Sons, Inc.}
\newblock


\bibitem[Medsker and Jain(2001)]%
        {medsker2001recurrent}
\bibfield{author}{\bibinfo{person}{Larry~R Medsker} {and} \bibinfo{person}{LC Jain}.} \bibinfo{year}{2001}\natexlab{}.
\newblock \showarticletitle{Recurrent neural networks}.
\newblock \bibinfo{journal}{\emph{Design and Applications}}  \bibinfo{volume}{5} (\bibinfo{year}{2001}), \bibinfo{pages}{64--67}.
\newblock


\bibitem[Meduri et~al\mbox{.}(2021)]%
        {meduri2021evaluation}
\bibfield{author}{\bibinfo{person}{Venkata~Vamsikrishna Meduri}, \bibinfo{person}{Kanchan Chowdhury}, {and} \bibinfo{person}{Mohamed Sarwat}.} \bibinfo{year}{2021}\natexlab{}.
\newblock \showarticletitle{Evaluation of machine learning algorithms in predicting the next SQL query from the future}.
\newblock \bibinfo{journal}{\emph{ACM Transactions on Database Systems (TODS)}} \bibinfo{volume}{46}, \bibinfo{number}{1} (\bibinfo{year}{2021}), \bibinfo{pages}{1--46}.
\newblock


\bibitem[Oppenheim(1999)]%
        {oppenheim1999discrete}
\bibfield{author}{\bibinfo{person}{A.V. Oppenheim}.} \bibinfo{year}{1999}\natexlab{}.
\newblock \bibinfo{booktitle}{\emph{Discrete-Time Signal Processing}}.
\newblock \bibinfo{publisher}{Pearson Education}.
\newblock


\bibitem[Oracle(2006)]%
        {selfmanage}
\bibfield{author}{\bibinfo{person}{Oracle}.} \bibinfo{year}{2006}\natexlab{}.
\newblock \bibinfo{booktitle}{\emph{{Oracle Database 10g Release 2: The Self-Managing Database}}}.
\newblock \bibinfo{type}{{T}echnical {R}eport}. \bibinfo{institution}{Oracle}.
\newblock


\bibitem[Padmanabhan et~al\mbox{.}(2003)]%
        {mdc}
\bibfield{author}{\bibinfo{person}{Sriram Padmanabhan}, \bibinfo{person}{Bishwaranjan Bhattacharjee}, \bibinfo{person}{Tim Malkemus}, \bibinfo{person}{Leslie Cranston}, {and} \bibinfo{person}{Matthew Huras}.} \bibinfo{year}{2003}\natexlab{}.
\newblock \showarticletitle{Multi-Dimensional Clustering: A New Data Layout Scheme in DB2}. In \bibinfo{booktitle}{\emph{SIGMOD '03}}. \bibinfo{pages}{637–641}.
\newblock


\bibitem[Pavlo et~al\mbox{.}(2017)]%
        {Pavlocidr17}
\bibfield{author}{\bibinfo{person}{Andrew Pavlo}, \bibinfo{person}{Gustavo Angulo}, \bibinfo{person}{Joy Arulraj}, \bibinfo{person}{Haibin Lin}, \bibinfo{person}{Jiexi Lin}, \bibinfo{person}{Lin Ma}, \bibinfo{person}{Prashanth Menon}, \bibinfo{person}{Todd~C. Mowry}, \bibinfo{person}{Matthew Perron}, \bibinfo{person}{Ian Quah}, \bibinfo{person}{Siddharth Santurkar}, \bibinfo{person}{Anthony Tomasic}, \bibinfo{person}{Skye Toor}, \bibinfo{person}{Dana~Van Aken}, \bibinfo{person}{Ziqi Wang}, \bibinfo{person}{Yingjun Wu}, \bibinfo{person}{Ran Xian}, {and} \bibinfo{person}{Tieying Zhang}.} \bibinfo{year}{2017}\natexlab{}.
\newblock \showarticletitle{Self-Driving Database Management Systems}. In \bibinfo{booktitle}{\emph{CIDR}}.
\newblock


\bibitem[Pedregosa et~al\mbox{.}(2011)]%
        {Scikit-learn}
\bibfield{author}{\bibinfo{person}{Fabian Pedregosa}, \bibinfo{person}{Ga\"{e}l Varoquaux}, \bibinfo{person}{Alexandre Gramfort}, \bibinfo{person}{Vincent Michel}, \bibinfo{person}{Bertrand Thirion}, \bibinfo{person}{Olivier Grisel}, \bibinfo{person}{Mathieu Blondel}, \bibinfo{person}{Peter Prettenhofer}, \bibinfo{person}{Ron Weiss}, \bibinfo{person}{Vincent Dubourg}, \bibinfo{person}{Jake Vanderplas}, \bibinfo{person}{Alexandre Passos}, \bibinfo{person}{David Cournapeau}, \bibinfo{person}{Matthieu Brucher}, \bibinfo{person}{Matthieu Perrot}, {and} \bibinfo{person}{\'{E}douard Duchesnay}.} \bibinfo{year}{2011}\natexlab{}.
\newblock \showarticletitle{Scikit-Learn: Machine Learning in Python}.
\newblock \bibinfo{journal}{\emph{J. Mach. Learn. Res.}} \bibinfo{volume}{12}, \bibinfo{number}{null} (\bibinfo{date}{nov} \bibinfo{year}{2011}), \bibinfo{pages}{2825–2830}.
\newblock
\showISSN{1532-4435}


\bibitem[Pincus(1991)]%
        {pincus1991approximate}
\bibfield{author}{\bibinfo{person}{Steven~M Pincus}.} \bibinfo{year}{1991}\natexlab{}.
\newblock \showarticletitle{Approximate entropy as a measure of system complexity.}
\newblock \bibinfo{journal}{\emph{Proceedings of the National Academy of Sciences}} \bibinfo{volume}{88}, \bibinfo{number}{6} (\bibinfo{year}{1991}), \bibinfo{pages}{2297--2301}.
\newblock


\bibitem[Radford et~al\mbox{.}(2019)]%
        {radford2019language}
\bibfield{author}{\bibinfo{person}{Alec Radford}, \bibinfo{person}{Jeffrey Wu}, \bibinfo{person}{Rewon Child}, \bibinfo{person}{David Luan}, \bibinfo{person}{Dario Amodei}, \bibinfo{person}{Ilya Sutskever}, {et~al\mbox{.}}} \bibinfo{year}{2019}\natexlab{}.
\newblock \showarticletitle{Language models are unsupervised multitask learners}.
\newblock \bibinfo{journal}{\emph{OpenAI blog}} \bibinfo{volume}{1}, \bibinfo{number}{8} (\bibinfo{year}{2019}), \bibinfo{pages}{9}.
\newblock


\bibitem[Sagi and Rokach(2018)]%
        {sagi2018ensemble}
\bibfield{author}{\bibinfo{person}{Omer Sagi} {and} \bibinfo{person}{Lior Rokach}.} \bibinfo{year}{2018}\natexlab{}.
\newblock \showarticletitle{Ensemble learning: A survey}.
\newblock \bibinfo{journal}{\emph{Wiley Interdisciplinary Reviews: Data Mining and Knowledge Discovery}} \bibinfo{volume}{8}, \bibinfo{number}{4} (\bibinfo{year}{2018}), \bibinfo{pages}{e1249}.
\newblock


\bibitem[Siddiqui et~al\mbox{.}(2020)]%
        {costmodel}
\bibfield{author}{\bibinfo{person}{Tarique Siddiqui}, \bibinfo{person}{Alekh Jindal}, \bibinfo{person}{Shi Qiao}, \bibinfo{person}{Hiren Patel}, {and} \bibinfo{person}{Wangchao Le}.} \bibinfo{year}{2020}\natexlab{}.
\newblock \showarticletitle{Cost Models for Big Data Query Processing: Learning, Retrofitting, and Our Findings}. In \bibinfo{booktitle}{\emph{SIGMOD}}. \bibinfo{pages}{99–113}.
\newblock


\bibitem[Taft et~al\mbox{.}(2018)]%
        {p-store}
\bibfield{author}{\bibinfo{person}{Rebecca Taft}, \bibinfo{person}{Nosayba El-Sayed}, \bibinfo{person}{Marco Serafini}, \bibinfo{person}{Yu Lu}, \bibinfo{person}{Ashraf Aboulnaga}, \bibinfo{person}{Michael Stonebraker}, \bibinfo{person}{Ricardo Mayerhofer}, {and} \bibinfo{person}{Francisco Andrade}.} \bibinfo{year}{2018}\natexlab{}.
\newblock \showarticletitle{P-Store: An Elastic Database System with Predictive Provisioning}. In \bibinfo{booktitle}{\emph{SIGMOD '18}} (Houston, TX, USA). \bibinfo{pages}{205–219}.
\newblock


\bibitem[Tang et~al\mbox{.}(2020)]%
        {CrocodileDB}
\bibfield{author}{\bibinfo{person}{Dixin Tang}, \bibinfo{person}{Zechao Shang}, \bibinfo{person}{Aaron~J. Elmore}, \bibinfo{person}{Sanjay Krishnan}, {and} \bibinfo{person}{Michael~J. Franklin}.} \bibinfo{year}{2020}\natexlab{}.
\newblock \showarticletitle{CrocodileDB in Action: Resource-Efficient Query Execution by Exploiting Time Slackness}.
\newblock \bibinfo{journal}{\emph{Proc. VLDB Endow.}} \bibinfo{volume}{13}, \bibinfo{number}{12} (\bibinfo{date}{aug} \bibinfo{year}{2020}), \bibinfo{pages}{2937–2940}.
\newblock


\bibitem[Taylor and Letham(2018)]%
        {taylor2018forecasting}
\bibfield{author}{\bibinfo{person}{Sean~J Taylor} {and} \bibinfo{person}{Benjamin Letham}.} \bibinfo{year}{2018}\natexlab{}.
\newblock \showarticletitle{Forecasting at scale}.
\newblock \bibinfo{journal}{\emph{The American Statistician}} \bibinfo{volume}{72}, \bibinfo{number}{1} (\bibinfo{year}{2018}), \bibinfo{pages}{37--45}.
\newblock


\bibitem[Vaswani et~al\mbox{.}(2017)]%
        {vaswani2017attention}
\bibfield{author}{\bibinfo{person}{Ashish Vaswani}, \bibinfo{person}{Noam Shazeer}, \bibinfo{person}{Niki Parmar}, \bibinfo{person}{Jakob Uszkoreit}, \bibinfo{person}{Llion Jones}, \bibinfo{person}{Aidan~N Gomez}, \bibinfo{person}{{\L}ukasz Kaiser}, {and} \bibinfo{person}{Illia Polosukhin}.} \bibinfo{year}{2017}\natexlab{}.
\newblock \showarticletitle{Attention is all you need}.
\newblock \bibinfo{journal}{\emph{Advances in neural information processing systems}}  \bibinfo{volume}{30} (\bibinfo{year}{2017}).
\newblock


\bibitem[Watkins and Dayan(1992)]%
        {watkins1992q}
\bibfield{author}{\bibinfo{person}{Christopher~JCH Watkins} {and} \bibinfo{person}{Peter Dayan}.} \bibinfo{year}{1992}\natexlab{}.
\newblock \showarticletitle{Q-learning}.
\newblock \bibinfo{journal}{\emph{Machine learning}}  \bibinfo{volume}{8} (\bibinfo{year}{1992}), \bibinfo{pages}{279--292}.
\newblock


\bibitem[Weinberger et~al\mbox{.}(2009)]%
        {featurehashing}
\bibfield{author}{\bibinfo{person}{Kilian Weinberger}, \bibinfo{person}{Anirban Dasgupta}, \bibinfo{person}{John Langford}, \bibinfo{person}{Alex Smola}, {and} \bibinfo{person}{Josh Attenberg}.} \bibinfo{year}{2009}\natexlab{}.
\newblock \showarticletitle{Feature Hashing for Large Scale Multitask Learning}. In \bibinfo{booktitle}{\emph{ICML '09}}. \bibinfo{pages}{1113–1120}.
\newblock


\bibitem[Wu et~al\mbox{.}(2018)]%
        {cardlearner}
\bibfield{author}{\bibinfo{person}{Chenggang Wu}, \bibinfo{person}{Alekh Jindal}, \bibinfo{person}{Saeed Amizadeh}, \bibinfo{person}{Hiren Patel}, \bibinfo{person}{Wangchao Le}, \bibinfo{person}{Shi Qiao}, {and} \bibinfo{person}{Sriram Rao}.} \bibinfo{year}{2018}\natexlab{}.
\newblock \showarticletitle{Towards a Learning Optimizer for Shared Clouds}.
\newblock \bibinfo{journal}{\emph{PVLDB}} \bibinfo{volume}{12}, \bibinfo{number}{3} (\bibinfo{date}{nov} \bibinfo{year}{2018}), \bibinfo{pages}{210–222}.
\newblock


\bibitem[Yuan et~al\mbox{.}(2020)]%
        {alibaba}
\bibfield{author}{\bibinfo{person}{Haitao Yuan}, \bibinfo{person}{Guoliang Li}, \bibinfo{person}{Ling Feng}, \bibinfo{person}{Ji Sun}, {and} \bibinfo{person}{Yue Han}.} \bibinfo{year}{2020}\natexlab{}.
\newblock \showarticletitle{Automatic view generation with deep learning and reinforcement learning}. In \bibinfo{booktitle}{\emph{{ICDE}}}. \bibinfo{pages}{1501--1512}.
\newblock


\bibitem[Zilio et~al\mbox{.}(2004)]%
        {db2advisor}
\bibfield{author}{\bibinfo{person}{Daniel~C. Zilio}, \bibinfo{person}{Jun Rao}, \bibinfo{person}{Sam Lightstone}, \bibinfo{person}{Guy Lohman}, \bibinfo{person}{Adam Storm}, \bibinfo{person}{Christian Garcia-Arellano}, {and} \bibinfo{person}{Scott Fadden}.} \bibinfo{year}{2004}\natexlab{}.
\newblock \showarticletitle{DB2 Design Advisor: Integrated Automatic Physical Database Design}. In \bibinfo{booktitle}{\emph{VLDB '04}}. \bibinfo{pages}{1087–1097}.
\newblock


\end{thebibliography}


\end{document}